\documentclass{aa} 

\usepackage{graphicx}
\usepackage[version=4]{mhchem}
\usepackage{color}
\usepackage{longtable}
\usepackage{txfonts}
\begin{document}

   \title{The UMIST database for astrochemistry 2022}

%   \subtitle{I. Overviewing the $\kappa$-mechanism}

   \author{T. J. Millar\inst{1}\thanks{email: tom.millar@qub.ac.uk}
          \and C. Walsh\inst{2}
          \and M. Van de Sande\inst{2,3}
          \and
          A. J. Markwick\inst{1,4}
          }

   \institute{Astrophysics Research Centre, School of Mathematics and Physics,
             Queen's University Belfast, University Road, Belfast BT7 1NN, UK
              \and
             Department of Physics and Astronomy, University of Leeds, Leeds LS2 9JT, UK
             \and 
             Leiden Observatory, Leiden University, P.O. Box 9513,
             2300 RA Leiden, The Netherlands
             \and
              Department of Physics and Astronomy, School of Natural Sciences, The University of Manchester, Manchester M13 9PL, UK}

   \date{Received 17 May 2023 / Accepted 27 October 2023}

% \abstract{}{}{}{}{} 
% 5 {} token are mandatory
 
  \abstract
  % context heading (optional)
  % {} leave it empty if necessary  
   {Detailed astrochemical models are a key component to interpret the observations of 
   interstellar and circumstellar molecules since they allow important physical properties 
   of the gas and its evolutionary history to be deduced.} 
  % aims heading (mandatory)
   {We update one of the most widely used astrochemical databases to reflect advances 
   in experimental and theoretical estimates of rate coefficients and to respond to the 
   large increase in the number of molecules detected in space since our last release in 2013.}
   % methods heading (mandatory)
  {We present the sixth release of the UMIST Database for Astrochemistry (UDfA), a major 
  expansion of the gas-phase chemistry that describes the synthesis of interstellar and 
  circumstellar molecules. Since our last release, we have undertaken a major review of 
  the literature which has increased the number of reactions by over 40\% to a total of 
  8767 and increased the number of species by over 55\% to 737. We have made a particular 
  attempt to include many of the new species detected in space over the past decade, 
  including those from the QUIJOTE and GOTHAM surveys, as well as providing references to the 
  original data sources.} 
  % results heading (mandatory)
   {We use the database to investigate the gas-phase chemistries appropriate to O-rich 
   and C-rich conditions in \object{TMC-1} and to the circumstellar envelope of the C-rich 
   AGB star \object{IRC+10216} and identify successes and failures of gas-phase only models.}
  % conclusions heading (optional), leave it empty if necessary 
   {This update is a significant improvement to the UDfA database.  For the dark 
   cloud and C-rich circumstellar envelope models, calculations match around 60\% of 
   the abundances of observed species to within an order of magnitude. There are a 
   number of detected species, however, that are not included in the model either 
   because their gas-phase chemistry is unknown or because they are likely formed 
   via surface reactions on icy grains. Future laboratory and theoretical work is 
   needed to include such species in reaction networks.}


   \maketitle
%
%-------------------------------------------------------------------

\section{Introduction}
\label{sec:intro}

   The first release of the UMIST Database for Astrochemistry (UDfA) was made public in 1991 \citep{mil91}.  It was, in main part, motivated by the various astrophysical applications - dark clouds, hot cores, circumstellar clouds, novae, and supernovae - that were being studied by the UMIST (University of Manchester Institute of Science and Technology)  Astrochemistry Group and the recognition that members of the group should have access to one set of reactions and rate coefficients that encompassed their various needs.  
   Once we had done this, it became clear that releasing it to the wider community would encourage the spread of astrochemical modelling as a tool both to interpret and to understand and make predictions for observations of molecules in space.

   It has been known for some time that gas-phase synthesis dominates the formation of many important interstellar molecules, such as \ce{CO}, \ce{N2}, \ce{HCO+}, \ce{N2H+}, \ce{H2D+}, and the unsaturated hydrocarbon chains. 
   For others, however, a detailed description of interstellar chemistry has to be augmented through reactions within and on the icy mantles of interstellar dust grains.  
   Gas-phase reactions do remain, however, the foundation on which such ice chemistry rests. It provides the feedstock for the grain mantle and can chemically process material that is removed from ices.  The detections of around 100 new molecules in space in the last few years have led to the addition of many new reactions and species to gas-phase networks and reflects the importance of chemical networks in the interpretation of molecular line observations. 
   This is particularly true in the current paradigm in which `bottom-up' synthesis determines the abundances of larger species.  Readers wishing to keep abreast of the latest detections of molecules in space should consult the Cologne Database for Molecular Spectroscopy\footnote[1]{https://cdms.astro.uni-koeln.de/} \citep{end16} or the Astrochemyst website\footnote[2]{https://www.astrochymist.org/} managed by David Woon.
   
   It is in this context that we present the sixth release of the UDfA, {\sc Rate22} (previous releases: {\sc Rate91} -- Millar et al. \citeyear{mil91}; {\sc Rate95} -- Millar et al. \citeyear{mil97}; {\sc Rate99} -- Le Teuff et al. \citeyear{let00}; {\sc Rate06} -- Woodall et al. \citeyear{woo07}; {\sc Rate12} -- McElroy et al. \citeyear{mce13}). The year suffix refers to the date at which we stopped collecting new data. Thus, {\sc Rate22} contains data published or in press up to the end of 2022.   Our fifth release \citep{mce13} contained 6173 reactions among 467 species and 13 elements and was developed in the context of the astronomical identification of around 150 molecular species at that time\footnote[3]{http://udfa.net/} and has proven a very popular source of astrochemical data with over 600 references to date. Applications of the data occur to sources as diverse as protoplanetary disks \citep{wal15}, infrared dark clouds \citep{ent22}, brown dwarf disks \citep{gre17}, the dynamical and chemical evolution of prestellar cores \citep{pri23}, photodissociation regions \citep{rol22}, external galaxies \citep{shi20}, and  cometary comae \citep{her17}. Its data have also been used in non-astrophysical applications, most commonly in plasma physics.

\section{The {\sc Rate22} database}
\label{sec:rate22}

\subsection{Description of the data}
\label{sec:data}

Our basic gas-phase ratefile, {\sc Rate22}\footnote[4]{https://umistdatabase.net}, now contains some 8767 individual rate coefficients. These correspond to 737 species involving 17 elements, increases of over 40\% and 55\% in reactions and species, respectively, from {\sc Rate12}. 
The additional elements are \ce{Al}, \ce{Ar}, \ce{Ca}, and \ce{Ti}. The basic format is that each line of data consists of 18 colon-separated entries: the first two are the reaction number and reaction type, defined in Table~\ref{tab:reac_type}, followed by two reactants and up to four products. 
The ninth entry denotes the number of temperature ranges, NTR, over which the rate coefficient is defined, while entries 10-12 give the values of $\alpha$, $\beta$ and $\gamma$ used to calculate the rate coefficients. 
Entries 13-14 give the temperature range over which the rate coefficient is defined, entry 15 provides the method by which the rate coefficient has been determined (M: measured; C: calculated; E: estimated; L: literature). Literature values have generally been harvested from reaction networks published within papers or in their supplementary materials. It also refers to rate coefficients for which we have no information on the method by which they have been determined. In some cases, it describes data taken from ratefiles sent to us on request by their authors. An example of the latter is the OSU high-temperature network \citep{har10}. We note, however, that the difference between estimated and literature values has been rather eroded over time. Of those reactions labelled `E', some 1501 are mutual neutralisation reactions, all of which are given the same rate coefficient \citep{har08, loo16}. Entry 16 is an estimate of accuracy, entry 17 gives, where available, some 7464 Digital Object Identifiers (DOI) or web page (URL) links that will take the user to the original data source. As part of this update, we have calculated the rate coefficients of 788 ion-dipole reactions using the approach described in Sect.~\ref{sec:newmols}. Of these, 521 currently contain no reference label; we shall update these DOI's upon publication of this paper which is the source of these rate coefficients. Entry 18 lists additional notes in which `Millar i-d' identifies these 788 ion-dipole reactions. A further 78 reactions identified by `Millar' refer to rate coefficients calculated or estimated by us. Of our total set of reactions, only 175 lack both a reference and an associated note.
Further details on the file format are given in \citet{mce13}.  Table~\ref{tab:reac_type} lists the different reaction types and compares the numbers of each in the {\sc Rate12} and {\sc Rate22} databases. In addition, it gives data on the origin and methodology by which the rate coefficients have been determined.

\begin{table}[ht]
\centering
\caption{Code, reaction type, and the number of each reaction type in {\sc Rate12} and {\sc Rate22}. Further information on the method by which the rate coefficient has been determined and the source of the data is given in Sect.~\ref{sec:data}.}   \label{tab:reac_type}
\begin{tabular}{clrr}
\hline \hline
Code & Reaction type & \multicolumn{2}{c}{Count} \\
\hline
     &                                  & R12   & R22 \\
\hline
AD   &   Associative Detachment         & 132   & 142 \\
CD   &   Collisional Dissociation       & 14    & 14 \\
CE   &   Charge Exchange                & 579   & 663 \\
CP   &   Cosmic-Ray Proton (CRP)        & 11    & {\bf 12} \\
CR   &   Cosmic-Ray Photon (CRPHOT)     & 249   & 428 \\
DR   &   Dissociative Recombination     & 531   & 831 \\
IN   &   Ion-Neutral                    & 2589  & 3447 \\
MN   &   Mutual Neutralisation          & 981   & 1501 \\
NN   &   Neutral-Neutral                & 619   & 1018 \\
PH   &   Photoprocess                   & 336   & 509 \\
RA   &   Radiative Association          & 92    & 152 \\
REA  &   Radiative Electron Attachment  & 24    & 30 \\
RR   &   Radiative Recombination        & 16    & 20 \\
\hline
     & Total reactions                  & 6173  & 8767 \\
     & Total species                    & 467   & 737\\
\hline
     & DOIs                             &       & 6990 \\
     & URLs                             &       & 474 \\
     & No references                    &       & 1303 \\
\hline
     & Measured                         &       & 1889 \\
     & Calculated                       &       & 1508 \\
     & Literature                       &       & 2996 \\
     & Estimated                        &       & 2374 \\
\hline
\end{tabular}
\end{table}

\subsection{Calculation of rate coefficients}
\label{sec:rates}

A specific rate coefficient is calculated from $\alpha$, $\beta$, and $\gamma$ according to its reaction type.  For binary reactions, this corresponds to the de Kooij-Arrhenius (KA), or modified Arrhenius, formula:
\begin{equation}
  k = \alpha\left(\frac{\mathrm{T}}{300}\right)^\beta \exp\left(\frac{-\gamma}{\mathrm{T}}\right) \quad \mathrm{cm^3s^{-1}},
\end{equation}
where T(K) is the gas temperature.  For cases in which a single rate coefficient is fit by different formulae over several contiguous temperature ranges (NTR $>$ 1), the entry is extended by including the appropriate data as additional colon-separated entries that have the same format as entries 10-18.  
For example, the rate coefficient of the ion-neutral reaction between \ce{H-} and \ce{CH2}, with NTR = 3, has been fitted to three KA formulae, one each for the temperature  ranges 30--100~K, 100--300~K and 300--1000~K \citep{yur20}.  In some complex cases, the rate coefficient of a particular reaction is best fitted by a sum of KA formulae over the same temperature range.  
In such cases we list the reaction as a series of independent rate coefficients each with its specific parameters. An example is the reaction between \ce{CN} and \ce{HC3N} to form \ce{NC4N}. A combined experimental and theoretical study of this reaction over the temperature range 5-400~K by \citet{che13} shows that the rate coefficient it is better fit as the sum of two KA formulae each of which is included as a separate line in the ratefile. 
In total, 10 reactions have complex rate coefficients described as the sum of two or more Arrhenius formulae so that the number of fully independent reactions in {\sc Rate22} is 8750.

For one-body reactions, that is those involving photons or cosmic ray particles. For cosmic-ray ionisation (CP), the formula for evaluation of the rate coefficients becomes
\begin{equation}
  k = \alpha \quad \mathrm{s^{-1}},
\end{equation}
where $\alpha$ is the cosmic-ray ionisation rate, while for those involving UV continuum photons (PH) the formula is:
\begin{equation}
  k = \alpha \exp(-\gamma A_V) \quad \mathrm{s^{-1}},
\end{equation}
where $\alpha$ represents the rate coefficient in the unshielded interstellar ultraviolet radiation field, $A_V$ is the dust extinction at visible wavelengths, and $\gamma$ is the parameter used to take into account the increased dust extinction at ultraviolet wavelengths. 
We have not explicitly included the effects of self-shielding in the database.  Such a process can occur in situations in which dissociating photons are absorbed through line rather than continuum processes and acts in addition to the extinction caused by dust grains. It can be important for molecules including \ce{H2}, \ce{CO}, \ce{N2}, \ce{OH}, and \ce{H2O} and depends on the column density of the molecule with these species showing almost complete shielding once a column density of around 10$^{15}$ cm$^{-2}$ is reached  \citep{hea17}. Numerical approaches often involve the use of look-up tables.  Details on the self-shielding of \ce{H2}, \ce{CO}, and \ce{N2} are discussed by \citet{ste14}, \citet{vis09}, \citet{li13} and \citet{hea14}.  The inclusion of self-shielding is particularly important in the atomic to molecular transition regions in molecular clouds, in photon-dominated regions (PDRs) circumstellar envelopes, and in certain regions of protoplanetary disks.

For cosmic ray-induced photoreactions (CR) the rate coefficient becomes:
\begin{equation}
 \label{eqn:crphot}
  k = \alpha\left(\frac{\mathrm{T}}{{300}}\right)^\beta \frac{\gamma}{1-\omega} \quad \mathrm{s^{-1}},
\end{equation}
where $\alpha$ is the cosmic-ray ionisation rate, $\gamma$ is the efficiency factor as defined in equation 8 of \citet{gre89}, and $\omega$ is the dust-grain albedo in the far ultraviolet, typically 0.4--0.6 at 150~nm for particles large compared to the wavelength and close to zero for very small grains \citep{jon13}.
We choose $\omega$ = 0.5 in our model calculations. The particular value of $\gamma$ depends on the Lyman band photons emitted by \ce{H2} following collisional excitation by the energetic electrons released in cosmic-ray ionisation. Thus the intensity and wavelength dependence of the cosmic ray-induced UV flux is very different from interstellar photons. Note that our $\gamma$ values are relative to the H$_2$ density and are different by a factor of two from those listed by \citet{hea17} to account for the factor 1/(1 - $\omega$) in Eqn.~\ref{eqn:crphot}. The cosmic-ray ionisation rates listed here are normalised to a total rate for electron production from cosmic ray ionisation (primarily from H$_2$ and He in dark clouds) of $\zeta_0$ = 1.36 $\times$ 10$^{-17}$ s$^{-1}$ \citep{pra80}. Rates for both direct cosmic ray ionisation  and cosmic ray-induced photoreactions can be scaled to other choices of the ionisation rate, $\zeta$, by multiplying the appropriate rate coefficients by $\zeta/\zeta_0$.

While the absolute rates for CR reactions are different from those of PH reactions for reasons mentioned above, we note that the number of CR reactions is significantly less than those of PH.  We have looked in detail at the reasons for this.  Three factors emerge: (i) ionisation by cosmic-ray photons is negligible for some species that readily ionised by the interstellar radiation field (see \citet{hea17} for examples); (ii) CR destruction of molecular ions, particularly where these react rapidly with \ce{H2}, is ignored; (iii) our literature searches showed that many hydrocarbon chain species have several product channels when subject to interstellar photodissociation but only one when cosmic-ray-induced photons are involved.  If we consider photorates for the species C$_n$H$_m$, n = 6-11, m = 0-2, there are 61 PH channels compared to 21 CR channels \citep{bet95, wak10, har10}.  Since it is not clear how the branching ratios might differ between PH and CR reactions, we have decided to retain, rather than augment, the original data.

\subsection{Caveats and limitations}
\label{sec:caveats}

Very recently, \citet{tin23} used quantum chemistry to investigate the thermochemistry of some 5768 reactions and the electronic energies of over 500 species contained in the KIDA database\footnote[5]{https://kida.astrochem-tools.org/} \citep{wak15a} identifying 306 that are endothermic.  
We have searched for each of these in our database finding that 53 overlap in terms of reactants and products with our list and should be removed. The fact that our number is so much smaller than in the KIDA database is due to a number of reasons: over 170 of the endothermic reactions identified in KIDA, particularly those involving large hydrocarbon chains, are not in {\sc Rate22}; {\sc Rate22} does not differentiate between certain isomers, for example between \ce{l-C3H} and \ce{c-C3H} nor between the cumulene \ce{H2C4} and diacetylene \ce{HC4H}, whereas KIDA does; energy barriers are already present in the {\sc Rate22} reactions; and the fact that {\sc Rate22} uses lower energy isomers where KIDA does not, for example \ce{SiOH+}, \ce{HOSO+}, \ce{HSCO+}, and \ce{H2CSH+} in {\sc Rate22} versus \ce{HSiO+}, \ce{HSO2+}, \ce{HOCS+}, and \ce{H3CS+} in KIDA. 

 We note too that many of the reactions excluded by \citet{tin23} are ion-neutral reactions with rate coefficients measured in low-pressure experiments to be a significant fraction of the collisional rate coefficient. 
 In these cases ionic products are normally identified by mass spectrometry so that a mis-identification of an ionic product may simply be due to the fact that the structures identified in the KIDA database are not the lowest energy isomer. 
 This is likely to be the case where independent measurements give similar products and rate coefficients. In other cases, particularly where the branching ratio of a channel is small at room temperature, an energy barrier may indeed be present.

As discussed in Sect.~\ref{sec:models}, the removal of these reactions does not make a significant difference to the comparison between observed and calculated abundances in both TMC-1 and IRC+10216. Both the full and the reduced ratefiles are made available to the community.

In addition to the differences between KIDA and {\sc Rate22} discussed above, and noting that there are significant overlaps in reactions and rate coefficients, there remain other important differences. 
The KIDA database includes several networks that incorporate both gas-phase and grain-surface reactions as well as several that describe chemistry in planetary atmospheres. 
Our new database includes reactions involving many of the complex organic molecules detected in recent years.  
It can also be applied to the circumstellar chemistries around both C-rich and O-rich AGB stars and will, no doubt, continue to be used in the range of applications listed in Sect.~\ref{sec:intro}.

As mentioned in Sect.~\ref{sec:intro}, many situations in astrochemistry need to involve surface or bulk chemistry on ice-mantled dust grains.  
The most important surface reaction is the association of two H atoms to form \ce{H2} on a grain surface. 
This reaction is not included explicitly in the {\sc Rate22} database. Our software does, however, include a term for this process in the ODEs describing the formation and destruction of H atoms. 
That is, it contains a term, HLOSS, which accounts for the loss of H atoms through the surface formation of \ce{H2}, the abundance of which is calculated, like that of electrons, by a conservation equation. 
The formation of \ce{H2} on grains has been the subject of many studies including the pioneering paper by \citet{hol71}.  
For example, \citet{caz04} described the process by which both physisorption and chemisorption contributed to \ce{H2} formation on bare silicate and carbonaceous surfaces (see the correction in \citet{caz10}), while \citet{leb12} discussed production in PDRs through both the Langmuir-Hinshelwood and Eley-Rideal mechanisms and \citet{fol18} investigated \ce{H2} formation on PAH particles. When using the database in other codes (that is, not the codes provided by us), care must to be taken to include grain-surface formation of \ce{H2}  (either implicitly or explicitly).

Other than \ce{H2} formation, grain surface chemistry is neglected here. It is an essential part of many astrochemical applications, particularly in the description of hot cores, hot corinos and protoplanetary disks.  
Many current models do incorporate such reactions but their networks have been developed mostly in isolation, that is, specific networks have been designed by individuals or groups for a particular purpose; there is no central source of `agreed' data. 
In particular, a quantitative description of surface chemistry needs to include parameters such as binding energies, diffusion barriers, reaction products and branching ratios, thermal and non-thermal desorption mechanisms, and the nature and composition of the ice, amongst others.  
The method of solving the grain chemistry ODEs is also a matter of choice, ranging from the computationally fast, but not always applicable, rate equation approach, to slower but more accurate Monte Carlo methods. \citet{cup17} discuss these issues in an extensive review article. 
Experimental and theoretical approaches that minimise uncertainties are underway but the lack of systematic results has prevented implementation of more general rules that might allow surface chemistry to be better described.  
We note, however, that the binding energies of some 133 molecules has been published by \citet{lig23}. 
Nonetheless, it has long been known experimentally that binding energies are dependent on the specific site at which a molecule sits within the ice \citep{col04}. 
More recent calculations by \citet{bol22}  and \citet{tin22} have provided the binding energies of over 20 molecules on model water ice clusters and show that all have a broad distribution of binding energies rather than a unique value.

As discussed above, values of the $\alpha$, $\beta$, and $\gamma$ parameters used to calculate rate coefficients are defined over specific temperature ranges.  
In many cases, these are fits to experimental data and it is appropriate to give a note of caution to their use at temperatures outside these ranges. 
\citet{rol11} noted that the extrapolation to low temperatures of rate coefficients in the {\sc Rate06} release could led to divergent and unphysical behaviour due to the incorporation of large negative $\gamma$ values.  
We corrected this and other issues noted by \citet{rol11} in our {\sc Rate12} release
and have been careful to ensure that our current rate coefficients are not divergent. 
The choice of whether or not to extrapolate a given rate coefficient outside of its given temperature range, particularly when determined experimentally, is not straightforward and it is unlikely that any one prescription works for all reactions. 
It is known, for example, that the quantum tunnelling of H atoms in reactions involving \ce{OH} can increase rate coefficients substantially at low temperatures \citep{hea18}. 
Choices also have to be made where rate coefficients are defined over two non-contiguous temperature ranges. 
For this situation we recommend that the rate coefficient is interpolated over the intermediate range, a solution that has the advantage that it does not allow a discontinuity in the value of the rate coefficient. \citet{rol11} provides an excellent summary of the issues involved in such cases. 

It is also possible that certain rate coefficients with large positive values of $\beta$ can diverge and be unphysical when extrapolated to temperatures much above the upper limit, T$_u$ of the range over which the fit is made (see \citet{sha23} for a discussion of this issue in {\sc Rate12}). 
This occurs only for NN reactions and we have identified 10 such reactions. 
There is no unique way to choose a formulation of these rate coefficients above T$_u$ and indeed, given the large values of $\beta$ involved in some, it can be dangerous to extrapolate to higher temperatures.
We therefore adopt a simple approach. For these reactions, we set NTR = 2, k(T > T$_u$) = k(T$_u$) and noted this fact in the individual entries. 

Finally, it is worth recalling that the number of reactions that describe the chemistry of individual species varies enormously from several hundred in some cases, for example  \ce{CO}, \ce{H2O}, \ce{C2H}, and \ce{C4H} amongst others, to a handful, particularly for many of the complex organic molecules and the large, newly detected, hydrocarbons. In the former case, one may expect that calculated abundances will be less sensitive to the values of rate coefficients. Particular circumstances can make this expectation fail. An example is diffuse cloud chemistry in which photodissociation dominates the loss of neutral species. In this case, their abundances should be inversely proportional to their unshielded photodissociation rates. In the latter case, the formation routes of large molecules are often unknown and have allocated rate coefficients that may be highly uncertain. 
The abundances of these species are likely to be almost directly proportional to the adopted rate coefficients although complex re-formation pathways may mitigate this behaviour to some extent. 
We plan a full sensitivity analysis of the {\sc Rate22} network for both dark clouds and circumstellar envelopes in order to identify critical reactions for further study.

\section{New and updated reactions and species}
\label{sec:newmols}

Here we present brief summaries of the chemistry associated with new species included since the {\sc Rate12} release as well as important updates to reaction rate coefficients.  
We have made a comprehensive literature search for species detected in space since our last release, identifying around 100 new molecules. 
We have searched for their formation routes in the literature and added destruction through a set of standard loss mechanisms: proton transfer reactions with species such as H$_3^+$, H$_3$O$^+$, HCO$^+$, and N$_2$H$^+$, and reactions with He$^+$ and C$^+$ which tend either to break molecular bonds or to build complexity. 
Charge exchange (CE) reactions with H$^+$, and to a lesser extent C$^+$, can also play an important role in chemistry. 
To help augment the formation routes of complex species, we have also added some mutual neutralisation (MN) reactions between common anions observed in IRC+10216 and TMC-1 -- \ce{C3N-}, \ce{C5N-}, \ce{C7N-}, \ce{C4H-}, \ce{C6H-}, and \ce{C8H-} -- and molecular cations (see \citet{mil07}, \citet{cor09} and \citet{wal09}). 
For these, we have adopted the rate coefficients suggested by \citet{har08}. Calculated abundances are particularly sensitive when a species has only one or a few known formation routes. 
Since this applies to many of the new molecules, we have allowed the dissociative recombination (DR) of their protonated forms to produce smaller fragments as well as the parent neutral. 
This approach is needed in dark clouds to prevent unphysical cycling of a species, that is, it prevents the effective loss rate of a molecule from being zero.  
In photon-dominated regions, photodissociation prevents such recycling.  We have included neutral-neutral (NN) reactions where we have evidence for these and destruction by interstellar and cosmic ray-induced photons. 
We have searched for permanent electric dipole moments and provide a list of over 270 values on our website. We then used these in an approximation of the Su-Chesnovich formula \citep{su82} to calculate ion-neutral (IN) rate coefficients:
    \begin{equation}
    \label{eqn:id}
    \mathrm{k} = 3.87 \times 10^{-9} \mu_\mathrm{D}\mu^{-1/2} (\mathrm{T}/300)^{-1/2} \quad \mathrm{cm^3 s^{-1}} \,, 
    \end{equation}
 where $\mu_D$ is the electric dipole moment of the neutral molecule in Debye and $\mu$ is the reduced mass of the reactants in atomic mass units. 
 This approximation is likely correct to within 20\% for species with dipole moments greater than 1 Debye and is well within the uncertainties of astrochemical model calculations. Where a specific rate coefficient has been measured at 300~K, we have used the measured rather than the calculated value to scale to other temperatures.

We have also taken the opportunity to relabel a small number of species in {\sc Rate12}. These are mostly protonated ions and have been relabelled to make clear their structural form. Table~\ref{tab:form} lists these.

\begin{table}[ht]
\centering
\caption{Relabelled species in {\sc Rate22}.}   \label{tab:form}
\begin{tabular}{llll}
\hline \hline
{\sc Rate12} &  {\sc Rate22} & {\sc Rate12} & {\sc Rate22}  \\
\hline
\ce{ H3CO+} &  \ce{H2COH+} & \ce{CH2CCH+} & \ce{H2CCCH+} \\
\ce{CH2CO+} & \ce{H2CCO+} & \ce{H3CS+} & \ce{H2CSH+} \\
\ce{C3H2O+} & \ce{HC3OH+} & \ce{H3C3O+} & \ce{HCCCHOH+} \\
\ce{COOCH4+} & \ce{HCOOCH3+} & \ce{HOCS+} & \ce{OCSH+} \\
\ce{HSiS+} & \ce{SiSH+} & \ce{H5C2O2+} & \ce{H2COOCH3+} \\
\ce{HSO2+} & \ce{HOSO+} & \ce{C7H5+} & \ce{CH3C6H2+} \\
\ce{H2C7N+} &\ce{HC7NH+} &\ce{H2C9N+} & \ce{HC9NH+} \\
\ce{HC4H} & \ce{C4H2} &. &  \\
\hline
\end{tabular}
\end{table}

 In the following, all fractional abundances are given relative to \ce{H2} unless otherwise stated. Within each sub-section, the discussion of the chemistry is ordered by the mass of the molecule.

\subsection{HCCNC, HCCNCH$^+$, HC$_4$NC, HC$_5$NH$^+$, C$_7$N$^-$ and HC$_{11}$N}
\label{sec:hc4nc}

\ce{HCCNC} and \ce{HCCNCH+} were first  detected in TMC-1 by \citet{kaw92} and \citet{agu22w}, respectively, at fractional abundances of 3 $\times$ 10$^{-10}$ \citep{cer20c} and 3 $\times$ 10$^{-12}$ \citep{agu22w} and  an abundance ratio \ce{HCCNC}/\ce{HNC3} = 5.8 \citep{cer20c}.  We note that \citet{vas18} have detected both \ce{HCCNC} and \ce{HNC3} in L1544 with an abundance ratio around 10.

The formation of \ce{HCCNC} is thought to be due to the DR of protonated \ce{HC3N}. We include five DR channels with rate coefficients taken from \citet{wak10} and adopt the same branching ratios for \ce{HCCNC} and \ce{HNC3}. Unfortunately, the experimental studies of the DR products of \ce{DC3ND+} do not distinguish between the \ce{DC3N} isomers \citep{vig12}. We include proton transfer reactions to form \ce{HCCNCH+} with rate coefficients calculated using Eqn.~\ref{eqn:id} as well as destruction involving reactions with \ce{C+} and \ce{He+}.  

Isocyanodiacetylene, \ce{HC4NC}, has been detected in TMC-1 by \citet{xue20} and \citet{cer20c} with the latter authors determining a fractional abundance of 3 $\times$ 10$^{-11}$, much less than the value of 1.8 $\times$ 10$^{-8}$ found for \ce{HC5N}.
Protonated cyanodiacetylene, \ce{HC5NH+}, has been tentatively detected in TMC-1 with a fractional abundance of 7.5 $\times$ 10$^{-11}$ by \citet{mar20j}.
 The basic chemistry is taken from Cernicharo et al. and, following these authors, we assume that the dissociative recombination of  \ce{HC5NH+} produces \ce{HC4NC} with 1\% efficiency. We have also included formation of \ce{HC4NC} through MN between \ce{HC4NC+} and the anions listed above. These particular processes may be important in both interstellar clouds and C-rich AGB outflows such as that around IRC+10216.

 \citet{cer23zt} have detected several transitions of the anion \ce{C7N-} in both TMC-1 and IRC+10216 allowing for accurate abundance estimates in both sources. It is a new species in {\sc Rate22} and we have included its formation through the reaction of \ce{H-} with \ce{HC7N} with a rate coefficient calculated by \citet{gia17} as well as through the radiative electron attachment between \ce{C7N} and \ce{e-} \citep{cor09}.  Loss occurs through photodetachment of the electron, reactions with H and O atoms as well as in several MN reactions with cations.

\ce{HC11N} is the largest cyanopolyyne detected to date and was identified in TMC-1 by \citet{loo21zd} with an abundance of 1.0 $\times$ 10$^{-10}$. It was not included in {\sc Rate12}. \citet{loo16} provide a detailed chemical model for the synthesis of \ce{HC11N} as well as several other large carbon-chain molecules, including \ce{HC7N} and \ce{HC9N}, \ce{CH3C7N}, and \ce{CH3C9N}, and radicals such as \ce{C10N} and \ce{C11N}. We have included close to 100 of their reactions in the database to enhance the formation of these large species.

\subsection{NC$_4$NH$^+$}
\label{nc4nh+}

Following the detection in TMC-1 of \ce{NCCNH+}, which is a tracer of the unobservable - at least through its rotational line emission - cyanogen molecule \ce{NCCN}, with an abundance 8.6 $\times$ 10$^{-12}$ by \citet{agu15q}, a related ion, protonated dicyanoacetylene, \ce{NC4NH+}, has been detected in the same source with a slightly smaller abundance 1.1 $\times$ 10$^{-12}$ by \citet{agu23zr}. 
 The chemistry of cyanogen and its related ions has been well studied in the laboratory and is described in detail in {\sc Rate12}.  Its larger counterpart \ce{NC4N} is likely to form through exothermic NN reactions involving \ce{C3N} and both HCN and HNC, studied in detail by \citet{pet04}. 
 A more efficient route though is the reaction of CN with \ce{HC3N} which has been studied experimentally over the temperature range 22--296~K by \cite{che13}. These authors also fit their data to higher temperature measurements and fit a double Arrhenius form from 5--400~K to the rate coefficient which has a value in excess of 10$^{-10}$ cm$^3$ s$^{-1}$ at 10~K.

\subsection{Propargyl and related species: CH$_2$CCH and CH$_2$C$_3$N}
\label{propargyl}

Propargyl, \ce{CH2CCH}, has been detected in TMC-1 with a fractional abundance of 10$^{-8}$ by \citet{agu21, agu22t} making it one of the most abundant hydrocarbon radicals yet detected. It was included in {\sc Rate12} but its ion-neutral reactions were erroneously given a T$^{-1/2}$ dependence. Since its dipole moment is small, 0.15~D, this dependence has been removed.

3-cyanopropargyl, \ce{CH2C3N}, has been detected in TMC-1 by \citet{cab21n} with a fractional abundance of 1.6 $\times$ 10$^{-11}$. This molecule is newly included in the {\sc Rate22} database and we follow the approach discussed by Cabezas et al. with formation driven by four NN reactions:

\begin{eqnarray}
   \mathrm{C} + \mathrm{CH_2CHCN} & \rightarrow & \mathrm{CH_2C_3N} + \mathrm{H} \\ 
   \mathrm{C_2} + \mathrm{CH_3CN} & \rightarrow & \mathrm{CH_2C_3N} + \mathrm{H} \\ 
    \mathrm{CN} + \mathrm{CH_2CCH} & \rightarrow & \mathrm{CH_2C_3N} + \mathrm{H} \\ 
   \mathrm{CH_3} + \mathrm{C_3N} & \rightarrow & \mathrm{CH_2C_3N} + \mathrm{H}, 
\end{eqnarray}
with rate coefficients as suggested by Cabezas et al..  Loss is assumed to be via the standard set of ion-neutral reactions with the DR of protonated \ce{CH2C3N} assumed to be divided equally between the product channels \ce{CH2C3N} + \ce{H} and \ce{CH2} + \ce{HC3N}.

\subsection{Methyl cyanopolyynes and their isomers}
\label{meth_cyanopoly}

\subsubsection{CH$_3$C$_3$N, HCCCH$_2$CN and H$_2$CCCHCN}
\label{sec:ch3c3n}

Two isomers of methyl cyanoacetylene, \ce{CH3C3N}, propargyl cyanide, \ce{HCCCH2CN}, and cyanoallene, \ce{H2CCCHCN}, have been detected in TMC-1 by \citet{mar21p}. All three species have very large dipole moments and we have calculated ion-neutral rate coefficients for new chemsirty involving these species using Eqn.~\ref{eqn:id}. We have, in the absence of evidence, assumed that the DR of the protonated forms of both propargyl cyanide and cyanoallene results in their parent neutral and propargyl, \ce{CH2CCH}, with a 50:50 branching ratio.
Formation of all three species  occurs primarily through neutral-neutral reactions involving CN: with propyne, \ce{CH3CCH}, and allene, \ce{CH2CCH2}, to form \ce{CH3C3N} \citep{abe15, car01}; with allene to form propargyl cyanide at a branching ratio of 0.1 \citep{bal02}, and with both propyne and allene to form cyanoallene, with branching ratios again taken from \citet{bal02}.

\subsubsection{CH$_3$C$_5$N and H$_2$CCCHC$_3$N}
\label{sec:ch3c5n}

\ce{CH3C5N} was detected in TMC-1 by \citet{sny06}. \citet{fue22zf}  used their most recent data to calculate its fractional abundance as 9.5 $\times$ 10$^{-12}$, about 8 times smaller than the value given by \citet{sny06}, arguing that their analysis fits better to the high-J transitions that they observe.  Its isomer, cyanoacetyleneallene \ce{H2CCCHC3N}, was detected in TMC-1 via the line stacking method by \citet{shi21} who derive a preferred value of its abundance as 2 $\times$ 10$^{-11}$.
\citet{fue22zf} subsequently confirmed this identification through the detection of 20 rotational transitions and derived an abundance of 1.2 $\times$ 10$^{-11}$.  

{\sc Rate12} assumed that the radiative association between \ce{CH3+} and \ce{HC5N} leads to \ce{CH3C5NH+} followed by DR to form \ce{CH3C5N} \citep{her89}. For its formation, we add the detailed pathways and rate coefficients suggested by \citet{shi21} and based on the laboratory study by \citet{fou14}  who studied the reaction of \ce{C3N} with allene and propyne down to 24~K.  The allene reactions are assumed to produce \ce{H2CCCHC3N} with 50\% efficiency. Shingledecker et al. assume that there are five product channels, with equal branching ratios, of the propyne reaction, including the formation of both \ce{CH3C5N} and \ce{H2CCCHC3N}. Two of these, however, are species not contained in the database, namely \ce{CH3CC} and \ce{HCCCH2C3N}. We thus substitute these species as their equivalents, propargyl \ce{CH2CCH}, and \ce{H2CCCHC3N}. In addition to these two reactions involving \ce{C3N}, we include  another neutral-neutral formation reaction, that between \ce{CN} and acetylene allene, \ce{H2CCCHCCH}.  Since the dipole moment of \ce{H2CCCHC3N} is large, 5.15~D (Cabezas, priv.comm.), we have also included the usual ion-neutral reactions with enhanced rate coefficients.

\subsubsection{CH$_3$C$_7$N and CH$_3$C$_9$N} 
\label{sec:ch3c7n}

\ce{CH3C7N} was discovered in TMC-1 by \citet{sie22v} who determined an abundance of 8.6 $\times$ 10$^{-12}$. 
Its chemistry in {\sc Rate12} is parallel to that of \ce{CH3C5N} but with the radiative association now involving \ce{HC7N} rather than \ce{HC5N} \citep{her89}.

\ce{CH3C9N} is newly included in the database and is as yet undetected in TMC-1.  Its formation is assumed to be through the radiative association of \ce{CH3+} and \ce{HC9N} with a rate coefficient assumed to be equal to that of the RA forming \ce{CH3C7NH+}.  It has a large dipole moment, 6.5~D, and we included the usual set of ion-neutral reactions with rate coefficients determined from Eqn.~\ref{eqn:id}.  \ce{CH3C9NH+} is formed in proton transfer reactions as well as in the reaction of \ce{N} with \ce{C10H5+} and is lost via DR and by reaction with \ce{C} atoms to form \ce{H2C11NH+} \citep{loo16}. For both species we have added some new ion-neutral loss reactions and calculated their rate coefficients using Eqn.~\ref{eqn:id}.

\subsection{CH$_2$CHCCH}
\label{sec:ch2chcch}

Vinyl acetylene was included in the {\sc Rate12} release but its chemistry was extremely limited with destruction only via interstellar and cosmic-ray-induced photons.  Here we have added formation through the measured \ce{C2H} + \ce{C2H4} reaction \citep{bou12}, destruction through IN and NN reactions involving \ce{C2H} to form ortho-benzyne, c-C$_6$H$_4$ \citep{zha11}, and CN to form, as suggested by \citet{lee21zi}, the more complex and newly-detected interstellar species vinylcyanoacetylene, \ce{H2CCHC3N}, and cyanovinylacetylene, \ce{HCCCHCHCN}, the chemistry of which are discussed further in Sect.~\ref{sec:vca}.

\subsection{Methyl polyacetylenes and their isomers}
\label{sec:meth_polyacet}

\subsubsection{CH$_3$C$_4$H and H$_2$CCCHCCH}
\label{sec:ch3c4h}

Allenyl acetylene, \ce{H2CCCHCCH}, was detected in TMC-1 by \citet{cer21s} with a fractional abundance of 1.2 $\times$ 10$^{-9}$, essentially identical to that of its isomer methyl diacetylene, \ce{CH3C4H}. Here we adopt the formation reactions suggested by Cernicharo et al., namely reactions of \ce{C2H} with \ce{CH2CCH2} and \ce{CH3CCH}, and of \ce{C2} with \ce{CH3CHCH2} which have been measured at low temperatures \citep{car01, dau08}. Product branching ratios for these two reactions have been determined by \citet{gou07}.   The dipole moment of allenyl acetylene is small, 0.63~D, so its ion-neutral reactions do not have a T$^{-1/2}$ dependence.

\subsubsection{CH$_3$C$_6$H and H$_2$CCCHC$_4$H}
\label{sec:ch3c6h}

Allenyl diacetylene, \ce{H2CCCHC4H}, an isomer of \ce{CH3C6H}, has been detected in TMC-1 with an abundance 2.2 $\times$ 10$^{-10}$ by \citet{fue22zf}  who also determine the abundance of \ce{CH3C6H} to be 7 $\times$ 10$^{-11}$. We adopt the same neutral-neutral formation routes as Fuentetaja et al. which are based on the low-temperature experiments of \citet{ber10}.  The {\sc Rate12} database contains many measured ion-neutral reactions that produce the \ce{C7H5+} ion.  Unfortunately, information on the isomeric forms of this ion produced in the experiments is unknown.  Here we assume that the isomeric forms are split equally between \ce{CH3C6H2+} and \ce{H2CCCHC4H2+}. Note that in {\sc Rate12} protonated \ce{CH3C6H} was labelled as \ce{C7H5+}.  We explicitly differentiate between the protonated forms of the two isomers in {\sc Rate22}.

\subsubsection{CH$_3$C$_8$H}
\label{sec:ch3c8h}

\citet{sie22v} searched unsuccessfully for \ce{CH3C8H} in TMC-1 deriving an upper limit to its abundance of 9.8 $\times$ 10$^{-10}$.  We have included it in {\sc Rate22} assuming that it is one of the products of the dissociative recombination of \ce{C9H5+} formed in the radiative association of \ce{C5H3+} and \ce{C4H2} \citep{her89}. The DR rate coefficient and its branching ratios are set equal to those of \ce{C7H5+}. The \ce{C9H5+} ion is also lost through reaction with \ce{N} atoms  leading eventually to \ce{HC9N} \citep{loo16}.

\subsection{Molecules containing the C-S bond}
\label{sec:s_chains}

\subsubsection{HCS and HSC}
\label{sec:hcs}

The first detections of the \ce{HCS} radical and its metastable isomer \ce{HSC} were made by \citet{agu18} in the dark cloud \object{L483}. \citet{cer21b} in their survey of S-bearing molecules in TMC-1 give an abundance of 5.5 $\times$ 10$^{-10}$ for \ce{HCS} and an upper limit of 1.3 $\times$ 10$^{-11}$ for \ce{HSC}.

We have added formation of \ce{HCS} via the fast reactions \ce{C} + \ce{H2S} \citep{dee06} and S with propargyl as well as some DR channels in the recombination of larger S-bearing ions. We note that {\sc Rate12} included photoionisation but not photodissociation of \ce{HCS}. We now include dissociation from both interstellar and cosmic ray-induced photons, setting the rates for these equal to those for \ce{HCO}.  The gas-phase synthesis of \ce{HSC} is essentially unknown (see \citet{agu18}) and we do not include it here.

\subsubsection{HCCS, HC$_2$S$^+$, H$_2$CCS, HC$_3$S, H$_2$CCCS, C$_4$S, HC$_4$S and C$_5$S}
\label{sec:s_hydrochains}

A number of new S-bearing hydrocarbons were discovered in TMC-1  by \citet{cer21b}, who discuss their chemistry in detail.  We have adopted their approach to the chemistry, taking many of the reactions and their rate coefficients from \citet{vid17} in their pioneering study of complex S-chemistry. Since all species have dipole moments greater than 1~D, we have re-calculated all ion-neutral reactions using Eqn.~\ref{eqn:id}, adding proton transfer and IN reactions involving abundant ions where these are missing.

\ce{HC2S+} has been detected in TMC-1 by \citet{cab22m} at an abundance of 1.1 $\times$ 10$^{-10}$. It appears in the {\sc Rate12} database but we have added a small number of reactions here. As \ce{C2S} has a very large proton affinity, 869.6 kJ mol$^{-1}$, larger than that of \ce{NH3} and smaller than that of \ce{C3S}, we have added some additional proton transfer reactions for the formation of \ce{HC2S+} and included proton transfer to \ce{C3S} as a loss mechanism.

The \ce{HC4S} molecule has been detected in TMC-1 with an abundance of 9.5 $\times$ 10$^{-12}$ by \citet{fue22zq}. We include the primary formation routes suggested by them, notably the reactions of \ce{S} and \ce{S+} with \ce{C4H3+} and \ce{C4H3}, respectively, and \ce{C} with \ce{H2CCCS}. To these we have added formation through \ce{S} and \ce{C4H2}. Loss reactions include those with \ce{N} atoms and charge exchange and proton transfer reactions with rate coefficients calculated using Eqn.~\ref{eqn:id}.

The thio-carbon chains, \ce{C4S} and \ce{C5S}, which have been detected in TMC-1 with fractional abundances of 3.8 $\times$ 10$^{-12}$ for \ce{C4S} and 5.0 $\times$ 10$^{-12}$ for \ce{C5S} by \citet{cer21b}, have large dipole moments and therefore the IN reactions involving both species have a T$^{-1/2}$ dependence at low temperature. \ce{C4S} was included in {\sc Rate12} with its formation dominated by MN reactions between anions and \ce{C4S+}. We have recalculated all IN rate coefficients using Eqn.~\ref{eqn:id} and added some neutral-neutral formation routes including the reactions between \ce{S} and \ce{C4H} and between \ce{C} and \ce{HC3S} \citep{vid17} as well as \ce{C} and  \ce{H2CCCS}. \ce{C5S} is a new species in the database and we have followed closely the approach for its formation and destruction suggested by Cernicharo et al. and Vidal et al., although we have again calculated IN rate coefficients using the ion-dipole approach. 

\subsubsection{NCS, HNCS and HSCN}
\label{sec:ncs}

Isothiocyanic acid, \ce{HNCS}, and its metastable isomer thiocyanic acid, \ce{HSCN}, were detected first in the \object{Sgr B2} molecular cloud by \citet{fre79} and \citet{hal09}, respectively. Subsequently, \citet{ada10} detected both species in TMC-1 while \citet{cer21b} detected \ce{NCS} in the same object and gave abundances for all three species, 7.8 $\times$ 10$^{-11}$, 3.8 $\times$ 10$^{-11}$ and 5.8 $\times$ 10$^{-11}$, respectively.   The chemistry of these three molecules are heavily interlinked and has been discussed in some detail by \citet{gro14}, who focused primarily on the routes forming \ce{HNCSH+} and \ce{H2NCS+}, and \citet{vid17}. We adopt their pathways but for consistency recalculate IN rate coefficients using Eqn.~\ref{eqn:id}.  

\subsubsection{HCSC$_2$H and HCSCN}
\label{sec:hcsc2h}

Propynethial, \ce{HCSC2H}, and cyanothioformaldehyde, \ce{HCSCN}, have been detected in TMC-1 by \citet{cer21f} with abundances of 3.2 $\times$ 10$^{-11}$ and 1.3 $\times$ 10$^{-10}$, respectively. For \ce{HCSC2H} we have included the formation reactions suggested by Cernicharo et al., namely \ce{S} with \ce{CH2CCH} and \ce{C2H} with \ce{H2CS}, with rate coefficients of 10$^{-10}$ cm$^3$ s$^{-1}$. We have included ion-neutral destruction with \ce{He+} and \ce{C+} as well as proton transfer reactions with the main proton carriers in dark clouds, \ce{H3+}, \ce{HCO+}, \ce{N2H+}, and \ce{H3O+}, followed by dissociative recombination with two equal product channels to \ce{HCSC2H} + \ce{H} and \ce{C2H2} + \ce{HCS}. To enhance its formation, we have also included a number of MN reactions as discussed in Sect.~\ref{sec:intro}. 

The formation of \ce{HCSCN} is treated in a similar manner to that of \ce{HCSC2H}, that is, we adopt two formation reactions \ce{S} + \ce{CH2CN} and \ce{CN} + \ce{H2CS}. These reactions, and those mentioned above for \ce{HCSC2H}, are analogues of the \ce{O} and \ce{H2CO} reactions that produce the interstellar molecules \ce{HCCCHO} and \ce{HCOCN} (see Sect.~\ref{sec:hcccho}).  

\subsubsection{CH$_3$SH and C$_2$H$_5$SH}
\label{sec:ch3sh}

Methyl mercaptan or methanethiol, \ce{CH3SH}, was first detected in the interstellar medium by  \citet{lin79} and has been seen in a wide variety of massive star-forming regions as well as in the low-mass star-forming system \object{IRAS 16293-2422} \citep{maj16}. It has not been detected in the classical cold dark clouds TMC-1 and L483. 
Two recent papers \citep{ela22, bou22} have performed extensive line surveys of two complex sources containing both hot and cold gas to unravel sulphur chemistry. Bouscasse at el. find \ce{CH3SH} only in hot gas surrounding  \object{G328.2551-0.5321} while el Akal et al. find it in the cold envelope of \object{Cyg X-N12} as well as in the hot gas. The formation of \ce{CH3SH} in the gas-phase appears to be difficult with no efficient routes identified that involve either ion-neutral or neutral-neutral reactions, leading to the conclusion that its chemistry is dominated by grain surface reactions (see, for example, \citet{maj16}).

One measured ion-neutral reaction that forms the C-S bond is:
\begin{equation}
\mathrm{S^+}  + \mathrm{CH_3CHO}  \rightarrow  \mathrm{CH_3SH^+} + \mathrm{CO}, 
\end{equation}
although the branching ratio to this channel is only 0.2 \citep{dec00}. We assume that the \ce{CH3SH+} ion is converted to \ce{CH3SH} through a series of MN reactions with anions.  Loss of \ce{CH3SH} occurs in ion-neutral reactions. Where there is no experimental evidence we have adopted the branching ratios suggested by Majumdar et al. although we have been more conservative in including reactions and products that are analogous to reactions involving \ce{CH3OH}.

Ethyl mercaptan, \ce{C2H5SH}, has been detected in its gauche (lowest energy) form toward the galactic centre by \citet{rod21b} who discuss its formation via possible grain-surface reactions (see also \citet{lam18}).  Due to the lack of feasible gas-phase formation routes, we have not included this species in the database.

\subsection{The hydrocarbon chains - C$_5$H$^+$, C$_5$H$_2$, C$_6$H$_2$, HCCCHCCC, C$_{10}$H and C$_{10}$H$^-$}
\label{sec:c5h+}

\ce{C5H+} was detected in TMC-1 by \citet{cer22g} at an abundance of 8.8 $\times$ 10$^{-12}$.  \ce{C5H+} has many formation routes but relatively few destruction reactions in {\sc Rate12}. We have added proton transfer reactions involving \ce{C5} with \ce{N2H+} and \ce{H3O+} to the former and included MN loss reactions with anions to the latter.

The linear cumulene pentatetraenylidene, \ce{C5H2}, has been detected with one of the lowest abundances seen in TMC-1, of 1.35 $\times$ 10$^{-12}$ \citep{cab21k}. Its chemistry was covered extensively in {\sc Rate12} but, as its dipole moment is 5.9~D, we have updated all IN rate coefficients involving this species using Eqn.~\ref{eqn:id}.

\citet{cab21k} also reported the abundances of the ortho- and para-cumulene forms of \ce{C6H2} with an ortho-para ratio of 3 and a total abundance equal to 8 $\times$ 10$^{-12}$ in TMC-1 while \citet{fue22zq} identified several emission lines in the same source from its isomer \ce{HCCCHCCC} with an abundance of 1.3 $\times$ 10$^{-11}$.  We do not include the latter species in the network.

The largest pure hydrocarbon chains detected to date are \ce{C10H} and \ce{C10H-} both of which have been observed, the former a tentative detection, by \citet{rem23zs} in TMC-1 with abundances of 2.0 $\times$ 10$^{-11}$ and 4.0 $\times$ 10$^{-11}$, respectively.  As noted by \citet{rem23zs}, this is the first neutral-anion pair in which the anion has a larger abundance than the neutral, a result difficult to explain with our current, and still limited, understanding of anion chemistry.  
These two species were already included in {\sc Rate12} but their chemistry was incomplete since, unlike smaller hydrocarbon chains, we did not allow their formation from larger chains.  We have rectified this by including \ce{C11H} and \ce{C11H-} in our species set and including the reaction \ce{O} + \ce{C11H-} to form \ce{C10H-} + \ce{CO}. 
The formation of a smaller hydrocarbon anion has been found in the measurements of \ce{O} with \ce{C2H-}, \ce{C4H-}, and \ce{C6H-} by \citet{eic07} who also note that the reactions become more exothermic as the chain length increases.  Although we fit the observed abundance of \ce{C10H} well, we find, in common with \citet{rem23zs}, that we underproduce \ce{C10H-} (see Sect.~\ref{sec:tmc1}).

\subsection{HCCCHO, c-C$_3$H$_2$O, H$_2$CCCO, HCOCN and CH$_2$CHCHO}
\label{sec:hcccho}

Propynal, \ce{HCCCHO}, was detected in TMC-1 by \citet{irv88} and subsequently observed by \citet{cer21f} who derived a fractional abundance of 1.5 $\times$ 10$^{-10}$. The chemistry of its isomers, c-C$_3$H$_2$O (cyclopropenone), with an abundance of 5.4 $\times$ 10$^{-12}$ \citep{loi16zl}, and \ce{H2CCCO} (propadienone), which has an upper limit of 1.1 $\times$ 10$^{-11}$ \citep{cer20d} have been discussed in some detail by Loison et al.  Here, we include only the chemistry of \ce{HCCCHO} and, in particular, the two reactions considered by \citet{cer21f}:
\begin{eqnarray}
   \mathrm{O} + \mathrm{CH_2CCH} & \rightarrow & \mathrm{HCCCHO} + \mathrm{H} \\
   \mathrm{C_2H} + \mathrm{H_2CO} & \rightarrow & \mathrm{HCCCHO} + \mathrm{H}, 
\end{eqnarray}
with rate coefficients of 10$^{-10}$ cm$^3$ s$^{-1}$. Ion-neutral rate coefficients are calculated using Eqn.~\ref{eqn:id} since the dipole moment of propynal is 2.78~D.

\citet{rem08} made the first detection of formyl cyanide (or cyanoformaldehyde), \ce{HCOCN}, in \object{Sgr B2(N)} and suggested that it was formed through the reaction of \ce{CN} with \ce{H2CO}, a suggestion strengthened by the quantum theoretical study of \citet{ton20}.   
More recently, however, \citet{wes23} studied the reaction experimentally over the range 32--103~K and performed high-level quantum calculations to find that the dominant product channel is \ce{HCN} + \ce{HCO}.  They found that the pathway to HCOCN formation is hindered by a submerged energy barrier. 
Based on a combination of experimental and theoretical approaches, they produced a fit to the rate coefficient over the temperature range 6--1500~K.  In {\sc Rate22}, the synthesis of \ce{HCOCN} is dominated by two NN reactions:
\begin{eqnarray}
   \mathrm{O} + \mathrm{CH_2CN} & \rightarrow & \mathrm{HCOCN} + \mathrm{H} \\
   \mathrm{CN} + \mathrm{CH_3CHO} & \rightarrow & \mathrm{HCOCN} + \mathrm{CH_3}, 
\end{eqnarray}
with rate coefficients equal to 10$^{-10}$ \citep{cer21f} and 6.5 $\times$ 10$^{-10}$ cm$^3$ s$^{-1}$ \citep{ton20}. In TMC-1, Cernicharo et al. find the fractional abundance of HCOCN is 3.5 $\times$ 10$^{-11}$.

The lowest energy conformer of propenal, \ce{CH2CHCHO}, was originally discovered by \citet{hol04} in the giant molecular cloud Sgr B2(N) and subsequently detected in TMC-1 by \citet{agu21zj} with an abundance of 2.2 $\times$ 10$^{-11}$.  It is formed through NN reactions \citep{tsa91, gou12, xie05}:
\begin{eqnarray}
   \mathrm{O} + \mathrm{C_3H_5} & \rightarrow & \mathrm{CH_2CHCHO} + \mathrm{H} \\
   \mathrm{CH} + \mathrm{CH_3CHO} & \rightarrow & \mathrm{CH_2CHCHO} + \mathrm{H} \\
    \mathrm{H_2CO} + \mathrm{C_2H_3} & \rightarrow & \mathrm{CH_2CHCHO} + \mathrm{H}.  
\end{eqnarray}    
We also include its formation through MN between the common anions and \ce{CH2CHCHO+}, formed in charge exchange with \ce{H+} and \ce{C+}.

\subsection{H$_2$NC and H$_2$CN}
\label{sec:h2nc}

\ce{H2CN} was included in {\sc Rate12}  but its chemistry was limited and dominated by neutral-neutral reactions taken from \citet{smi04}. Since its dipole moment is 2.54~D, we have added our usual suite of ion-neutral reactions with rate coefficients calculated by Eqn.~\ref{eqn:id}. \ce{H2NC}, the highest energy isomer of \ce{H2CN}, was discovered in the dark clouds L483 and \object{B1-b} by \citet{cab21zm} although they did not detect it nor \ce{H2CN} in TMC-1, with upper limits to their fractional abundances of 3.2 $\times$ 10$^{-11}$ and 4.8 $\times$ 10$^{-11}$, respectively. 
Although the energy difference is 29.9 kJ mol$^{-1}$, the abundance ratio of these two molecules is close to 1 in both cold clouds.  We have adopted a similar chemistry for \ce{H2NC}, adding the measured reaction between \ce{C} and \ce{NH3} \citep{bou15} to form both isomers with equal efficiency. Neutral atom destruction of \ce{H2NC} is assumed to proceed at the same rate as that of \ce{H2CN}.

\subsection{C$_\mathrm{n}$O and HC$_\mathrm{n}$O, n = 4-9}
\label{sec:cno_hcno}

In recent years, there has been a number of successful (and unsuccessful) observational searches in TMC-1 for long carbon-chain molecules terminated by an oxygen atom \citep{mcg17a,cor17zn, cer21u}. Between them, these papers give fractional abundances for species \ce{C2O} to \ce{C7O} and \ce{HCO} to \ce{HC7O}.

Two major mechanisms have been suggested to form these molecules. The first is based on experimental work on the ternary association reactions between CO and unsaturated hydrocarbon ions from which are deduced two-body radiative association rate coefficients at 10~K to form HC$_{\mathrm{n}+1}$O$^+$, HC$_{\mathrm{n}+1}$OH$^+$ ,and H$_2$C$_{\mathrm{n}+1}$OH$^+$ ions, n = 2-6 \citep{ada89}. We have adjusted their pre-exponential factor and adopted a (T/300)$^{-2.5}$ dependence so that the rates can be used at other temperatures. 
Such a dependence is based on the modified thermal approach \citep{har10} where the exponent depends on the sum of the rotational degrees of freedom of the reactants.  In this case we assume that the hydrocarbon ions are non-linear. If instead, the dominant collisional ion is linear, then the dependence should be (T/300)$^{-2.0}$.  
This would mean that we over-estimate the rate constant at 10~K by a factor of 5.5. In addition, we have extended the set of reactions to include the \ce{C9O} species.  Since the 10~K rates are already close to their collisional values for n = 5, we assume that the radiative association rates for these larger species are equal to that for the \ce{C5H+} + \ce{CO} association.

Once formed these ions can then undergo DR to form the range of neutrals observed. It should be clear, however, that the actual abundances of the (H)C$_\mathrm{n}$O species are particularly sensitive to the branching ratios for these DR processes. In the absence of any evidence, we have chosen a procedure that is easy to implement (and to change in the light of future information) - each ion formed through radiative association dissociates into two channels with equal branching ratios. Using \ce{C7O} as an example:
\begin{align}
   &\mathrm{HC_7O^+} + \mathrm{e^-} \rightarrow \mathrm{C_6} + \mathrm{CO}  &0.5  \\ 
   &\mathrm{HC_7O^+} + \mathrm{e^-} \rightarrow  \mathrm{C_7O} + \mathrm{H}  &0.5   \\
   &\mathrm{HC_7OH^+} + \mathrm{e^-} \rightarrow  \mathrm{C_7O} + \mathrm{H_2}   &0.5   \\ 
   &\mathrm{HC_7OH^+} + \mathrm{e^-} \rightarrow \mathrm{HC_7O} + \mathrm{H}  &0.5  \\
   &\mathrm{H_2C_7OH^+} + \mathrm{e^-} \rightarrow  \mathrm{C_7O} + \mathrm{H_2} + \mathrm{H}  &0.5  \\ 
   &\mathrm{H_2C_7OH^+} + \mathrm{e^-}  \rightarrow  \mathrm{HC_7O} + \mathrm{H_2}  &0.5. 
\end{align}

The second route to these long chain molecules has been proposed by \citet{cor12} and is based on the laboratory measurements of \citet{eic07}. These form the chain species through reactions involving \ce{O} atoms and hydrocarbon anions, such as:
\begin{eqnarray}
    \mathrm{O} + \mathrm{C_6H^-} & \rightarrow  & \mathrm{HC_6O} + \mathrm{e^-} \\
    \mathrm{O} + \mathrm{C_6H^-} & \rightarrow  & \mathrm{CO} + \mathrm{C_5H^-}, 
\end{eqnarray}
again with equal branching ratios.

In both of these scenarios, a certain simplicity of implementation is adopted.  The real situation is likely to be more complex with other possible product channels and branching ratios so our approach, in the absence of further investigation, likely overestimates the gas-phase formation of these neutral chain molecules.

The C$_\textrm{n}$O species all have large dipole moments \citep{moa95}; those of HC$_\textrm{n}$O are smaller \citep{moh05} but still larger than 1~D. We use Eqn.~\ref{eqn:id} to generate the rate coefficients for ion-neutral reactions.

\subsection{CH$_3$CHCHCN}
\label{sec:ch3chchcn}

The trans- and cis- forms of crotononitrile, \ce{CH3CHCHCN}, are two of five cyano derivatives of propene discovered in TMC-1 by \citet{cer22zo}, the others being methacrylonitrile, \ce{CH2CH(CH3)CN}, and gauche- and cis-allyl cyanide, \ce{CH2CHCH2CN}.  All species have very similar abundances $\sim$10$^{-11}$ indicating that they are likely to form in the same manner.  Cernicharo et al. note that the reaction of \ce{CN} with propene, \ce{CH3CHCH2}, has been measured to be fast down to 23~K \citep{mor10} with products that could include the above species as well as vinyl cyanide, \ce{CH2CHCN}, with a branching ratio that greatly favours vinyl cyanide.

Since the detailed chemistry of these cyano derivatives and the interplay between them is not understood, here we only include crotononitrile as a proxy for the others and allocate it a branching ratio of 30\%, with vinyl cyanide allocated 70\% in the \ce{CN}-\ce{CH3CHCH2} reaction.  
In addition to photodissociation, we include the usual ion-neutral loss reactions.  As noted by Cernicharo et al., this formation route is now a very inefficient mechanism for forming these cyano derivatives.  This is a result of a very much decreased production rate of propene following the study by \citet{lin13} who showed that the radiative associations of both \ce{C3H3+} and \ce{C3H5+} with \ce{H2} possess energy barriers and do not proceed at low temperatures.  
The result is a precipitous decrease in the abundance of \ce{C3H7+} and hence of propene calculated under dark cloud conditions, well below that observed, and as a result, leads to very low abundances for these cyano derivatives.

\subsection{Cyanates, their isomers and related species}

\subsubsection{HNCO and its isomers}
\label{sec:hnco_isomers}

Isocyanic acid, \ce{HNCO}, has three metastable isomers whose chemistry has been discussed in detail by \citet{qua10}.  Here we updated ion-neutral rate coefficients using Eqn.~\ref{eqn:id} and added proton transfer reactions with \ce{HCO+}, \ce{N2H+}, and \ce{H3O+}.

\subsubsection{OCN, H$_2$NCO$^+$ and H$_2$OCN$^+$}
\label{sec:ocn}

These molecules have not been observed as yet in TMC-1 but the first two were detected in L483 by \citet{mar18}. \ce{OCN} has a fairly extensive chemistry in the {\sc Rate12} database but the only proton transfer reaction included was with \ce{H3+} which erroneously had a T$^{-1/2}$ dependence. We have corrected this and included a wider range of such reactions. We have also added a new formation reaction \ce{CN} + \ce{O2} for which \citet{sim92} measured the rate coefficient from 26-295~K. At higher temperatures we adopt the rate coefficient recommended by \citet{bau94}.

The chemistry of \ce{H2NCO+} and \ce{H2OCN+} is very limited in {\sc Rate12} where they are formed only by proton transfer of \ce{H3+} with \ce{HNCO} and \ce{HOCN} respectively, and by the reactions of \ce{H2}  with \ce{HCNO+} and \ce{HOCN+}, respectively.  Marcelino et al. have shown that the \ce{H2} reactions are, however, endothermic. We have therefore removed the \ce{H2} reactions and added additional proton transfer reactions. Furthermore, the only loss routes for the two ions in {\sc Rate12} are via DR; we have added MN reactions as a result. 

\subsubsection{CH$_3$NCO, HOCH$_2$CN and C$_2$H$_5$NCO}
\label{sec:ch3nco}

The first detection of interstellar methyl isocyanate, \ce{CH3NCO}, was made in Sgr B2(N) by \citet{hal15}. \citet{maj18} have discussed its gas and grain chemistry. Their list of reactions contains only one gas-phase formation pathway, the reaction of \ce{HNCO} and \ce{CH3}. 
Their calculation of the rate coefficient, however, has an energy barrier of 8040~K making it unimportant in most interstellar sources, with the possible exception of shocked gas. Their modelled abundance of \ce{CH3NCO} therefore depends completely on grain surface reactions.  Subsequently, \citet{gor20} revised the value of the rate coefficient to a temperature-independent value of 1.0 $\times$ 10$^{-12}$ cm$^3$ s$^{-1}$. whereas Qu\'{e}nard et al. (2018) adopt a value of 5.0 $\times$ 10$^{-11}$ cm$^3$ s$^{-1}$.  Since this is the single gas-phase production reaction for \ce{CH3NCO}, its calculated abundance is directly proportional to the value of the rate coefficient.  Here we adopt the value of Qu\'{e}nard et al. to maximise its abundance but note that this will be over-estimated by a factor of 50 if Gorai et al.'s value is correct. We have taken the bulk of our additional gas-phase reactions from Majumdar et al., updating them using Eqn.~\ref{eqn:id} where appropriate.

Glycolonitrile, \ce{HOCH2CN}, an isomer of \ce{CH3NCO}, has been detected in two interstellar sources IRAS 16293-2422B \citep{zen19} and \object{Serpens SMM1} \citep{lig21}. There are, however, no gas-phase routes identified to form this molecule and we do not include it in the ratefile.

Finally, we note in passing that ethyl isocyanate, \ce{C2H5NCO}, has been detected in the galactic centre cloud \object{G+0.693-0.027} by \citet{rod21a}. There is no gas-phase chemistry available for the formation of this molecule and it is not included in the database.

\subsection{NH$_2$CHO}
\label{sec:nh2cho}

Formamide, \ce{NH2CHO}, is observed in many interstellar clouds although not in TMC-1, where its upper limit is 5 $\times$ 10$^{-12}$ \citep{cer20d}, and was not included in {\sc Rate12}, consistent with the subsequent findings by \citet{red14a, red14b} that potential gas-phase ion-neutral syntheses had energy barriers despite being exothermic. 
The molecule was added to the current database following the calculation by \citet{bar15} that the reaction between \ce{NH2} and \ce{H2CO} was efficient in making \ce{NH2CHO}.  However, a very recent experimental and theoretical study of this reaction down to 34~K shows that the channel producing \ce{NH2CHO} has an energy barrier of over 1800~K and therefore cannot act as a production route in interstellar clouds \citep{dou22}. 
There appears to be no route to \ce{NH2CHO+} thereby ruling out MN as a low-temperature route to formamide. We do, however, leave this species in the database since the high temperatures that need to be reached to overcome the energy barrier can be realised in hot, inner circumstellar envelopes (CSEs) and in post-shock gas.

\subsection{HCOOCH$_3$, CH$_3$COOH, HOCH$_2$CHO and (CHOH)$_2$}
\label{sec:hcooch3_isomers}

Acetic acid, \ce{CH3COOH}, glycolaldehyde \ce{HOCH2CHO}, and 1,2-ethenediol, \ce{(CHOH)2}, hereafter AA, GA, and ED, respectively, are isomers of the common interstellar molecule methyl formate, \ce{HCOOCH3}, hereafter MF, the only one of the four included in {\sc Rate12}. 
Its formation there, though, is very inefficient, essentially due to one slow ion-neutral reaction, that between \ce{CH3OH2+} and \ce{H2CO} to produce protonated MF. The {\sc Rate12} model produces a maximum abundance of less than 10$^{-16}$ under TMC-1 conditions (Sect.~\ref{sec:tmc1}).  MF is the only one of these four species to be detected in TMC-1 with an abundance of 1.1 $\times$ 10$^{-10}$ \citep{agu21zj}. The others are observed in molecular clouds associated with star forming regions.

\citet{asc19} have measured the rate coefficient of \ce{He+} with MF and re-evaluated other proton transfer reactions.  We have followed their suggestions but used Eqn.~\ref{eqn:id} to calculate total rate coefficients except where measured values are available.

Gas-phase formation of AA and GA, as currently understood, is an inefficient process and most models that attempt to reproduce their abundances in interstellar clouds tend to rely on grain surface chemistry. \citet{sko18} discuss this in some detail. Their gas-phase scheme begins with the hydrogen abstraction reactions of \ce{Cl} and \ce{OH}  with ethanol:

\begin{eqnarray}
       \mathrm{Cl} + \mathrm{C_2H_5OH} & \rightarrow  & \mathrm{CH_3CHOH} + \mathrm{HCl} \\
    \mathrm{Cl} + \mathrm{C_2H_5OH} & \rightarrow  & \mathrm{CH_2CH_2OH} + \mathrm{HCl}. 
\end{eqnarray}
These radicals then react with O atoms - \ce{CH3CHOH} to form AA and \ce{CH2CH2OH} to form GA, amongst other products.  Skouteris et al. calculate rate coefficients and branching ratios for all these reactions which we adopt here. As usual, we have used Eqn.~\ref{eqn:id} to calculate the rate coefficients of the ion-neutral loss reactions. We have taken DR rate coefficients for both ionised and protonated AA and GA from Walsh (priv. comm.) with some adjustments to branching ratios for channels involving products \ce{CH3CO} and \ce{COOH} that are not contained in this release.

(Z)-1,2-ethenediol, which we write in the database as \ce{HOCHCHOH}, has recently been detected in the Galactic Centre cloud \object{G+0.693-0.027} by \citet{riv22}. They find a low rotational excitation temperature of about 8.6~K, and an abundance ratio with respect to glycolaldehyde of 0.19. We include the two gas-phase routes suggested by Rivilla et al.:

 \begin{eqnarray}
       \mathrm{H_2CO} + \mathrm{CH_2OH} & \rightarrow  & \mathrm{HOCHCHOH} + \mathrm{H} \\
    \mathrm{CH_2CHOH} + \mathrm{OH} & \rightarrow  & \mathrm{HOCHCHOH} + \mathrm{H}, 
\end{eqnarray}  
with temperature-independent rate coefficients 2 $\times$ 10$^{-10}$ cm$^3$ s$^{-1}$.

\subsection{CH$_3$NH, CH$_2$NH$_2$ and CH$_3$NH$_2$}

Methylamine, \ce{CH3NH2}, was one of the earliest detected interstellar molecules having been observed in the Giant Molecular Clouds Sgr B2 and \object{Orion A} by \citet{kai74} and \citet{fou74}, respectively.  It was not included in earlier versions of the database. Its gas-phase formation is highly uncertain with the dominant  pathway initiated by the radiative association of \ce{CH3+} and \ce{NH3} to form \ce{CH3NH3+} \citep{her85} followed by dissociative recombination. 
It has recently been shown experimentally that it has a fast loss reaction with OH at 22~K in which the radicals \ce{CH2NH2} and \ce{CH3NH} are produced \citep{gon22} with a rate coefficient some 20 times larger than that at 300~K. \citet{puz20} have made a theoretical study of the H-atom abstraction reaction  of \ce{CH3NH2} with CN and we adopt their rate coefficients and branching ratios to \ce{CH2NH2} and \ce{CH3NH}.  These two radicals have also been added to {\sc Rate22}. 
We note that \citet{sch22} have recently obtained a laboratory rotational spectrum of protonated methylamine and searched for it in several cores in Sgr B2, determining an upper limit of $\sim$ 10$^{-10}$.  Finally, we note that \citet{wlo88} have studied the reaction of \ce{Si+} with \ce{CH3NH2} showing that it forms \ce{SiNH+} and hence \ce{SiN} following DR.

\subsection{Cyclic hydrocarbons -- c-C$_5$H, c-C$_3$HCCH, c-C$_5$H$_6$, c-C$_6$H$_4$, c-C$_6$H$_5$, c-C$_5$H$_4$CCH$_2$, 
c-C$_5$H$_5$CCH, c-C$_5$H$_5$CN, C$_6$H$_5$CCH, c-C$_6$H$_5$CN, c-C$_9$H$_8$ and c-C$_9$H$_7$CN }
\label{cyclic_hydro}

c-C$_5$H, written as \ce{C3CCH} to distinguish it from linear \ce{C5H} in the database, has been detected with an abundance of 9 $\times$ 10$^{-12}$ in TMC-1 \citep{cab22ze}. They suggest two formation reactions, one the reaction of \ce{C2H} with c-C$_3$H, which retains the triangular \ce{C3} structure in c-C$_5$H. 
The second is the reaction of \ce{C} atoms with diacetylene, a reaction they suggest may produce both linear and cyclic \ce{C5H}, although they note that production of the latter is close to thermo-neutral and may not proceed at low temperatures. We have therefore adopted their suggestion and taken a branching ratio of 10\% for c-C$_5$H.  In addition to destruction by N, C and O atoms, we have added our usual suite of dipole-enhanced IN reactions.

Ethynyl cyclopropenylidenene, c-C$_3$HCCH, written simply as \ce{C3HCCH} in the database, and cyclopentadiene, c-C$_5$H$_6$, and written \ce{C5H6}, were detected in TMC-1 with fractional abundances of 3.1 $\times$ 10$^{-11}$ and 1.2 $\times$ 10$^{-9}$, respectively, by \citet{cer21zp, cer21zb}. 
Since  the basic structure of the former species is a C$_3$-triangle, its chemistry is thought to be dominated by the reaction of \ce{C2H} with cyclopropenylidene, c-C$_3$H$_2$, given that both reactants are abundant in TMC-1 and that other reactions between \ce{C2H} and hydrocarbon radicals are known to be fast at low temperature. c-C$_3$HCCH has a large dipole moment, 3.54~D, so again we have used Eqn.~\ref{eqn:id} to calculate ion-neutral rate coefficients. Theoretical and/or experimental studies on reactions that might lead to this molecule are clearly needed.

The gas-phase chemistry of cyclopentadiene, c-C$_5$H$_6$, is also likely to be incomplete. Its neutral-neutral formation is entirely dependent on one reaction between \ce{CH} and butadiene, \ce{CH2CHCHCH2} \citep{cer21zb}.   
The dissociative recombination of \ce{C5H7+} can also produce it but, other than proton transfer reactions  of c-C$_5$H$_6$ forming \ce{C5H7+}, which are cyclic in nature and not therefore true formation routes to \ce{C5H6}, this ion is only produced in slow ion-molecule reactions involving \ce{C3H4+} and \ce{C3H5+}. 
In general, the unreactivity of large carbon-bearing species with \ce{H2} makes the synthesis of such molecules with more than 3-4 hydrogen atoms, and in particular \ce{C5H7+}, inefficient.

\citet{cer21y} detected ortho-benzyne, hereafter \ce{C6H4}, with a fractional abundance of 5.0 $\times$ 10$^{-11}$ in TMC-1. They provide an exhaustive discussion of a total of 14 possible formation routes, many of which possess energy barriers. 
Here, we include the four major neutral-neutral routes identified by Cernicharo et al.  The exothermic, barrierless reaction between \ce{C2H} and vinyl acetylene, \ce{CH2CHCCH}, has been measured by \citet{zha11} with an estimated rate coefficient of 8 $\times$ 10$^{-11}$ cm$^3$ s$^{-1}$. 
The other three syntheses involve the reactions of \ce{C2H4}, \ce{C3H}, and \ce{C3H2} with \ce{C4H}, \ce{CH3CCH}, and \ce{CH2CCH}, respectively. We have also included the reaction between \ce{C3H} and \ce{CH2CCH2} with all rate coefficients set to 10$^{-10}$ cm$^3$ s$^{-1}$. Loss reactions include charge exchange with \ce{H+} and \ce{C+} and proton transfer reactions. The phenyl radical, hereafter C$_6$H$_5$, has not yet been detected in the interstellar medium but plays an important role in the chemistry of the benzene-like ring molecules.

Fulvenallene, \ce{C5H4CCH2}, was detected with an abundance of 2.7 $\times$ 10$^{-10}$ in TMC-1 by \citet{cer22zk}. They suggested that it was formed as one of the products in the reaction of \ce{C2H} with cyclopentadiene, \ce{c-C5H6}, the other two channels producing 1- and 2-ethynyl cyclopentadiene, \ce{c-C5H5CCH}.  
Since both propene and \ce{c-C5H6} play an important role in the synthesis of fulvenallene we have added a number of ion-molecule reactions, taken from the compilations by  \citet{ani86} and \citet{ani93}, aimed at increasing the abundances of these species. These include:
\begin {eqnarray}
\mathrm{C_2H_5^+} + \mathrm{CH_3CH_3} & \rightarrow & \mathrm{C_3H_7^+} + \mathrm{CH_4} \\
\mathrm{C_3H_5^+} + \mathrm{C_2H_4} & \rightarrow & \mathrm{C_5H_7^+} + \mathrm{H_2} \\
\mathrm{CH_2CCH^+} + \mathrm{C_2H_4} & \rightarrow & \mathrm{C_5H_5^+} + \mathrm{H_2}. 
\end{eqnarray}

\citet{cer22zk} suggest that the radiative association reaction between \ce{l-C3H3+}, denoted \ce{CH2CCH+} in the database, and \ce{C2H4} forms \ce{C5H7+} with a very large rate coefficient.  Although this product ion was detected in the laboratory, the measurement was performed under the relatively high pressure conditions within an ion cyclotron resonance experiment, so that the rate coefficient determined is more likely to be the two-body equivalent of a three-body process rather than that of radiative association.

The two isomers 1- and 2-ethynyl cyclopentadiene, \ce{C5H5CCH}, have been detected in TMC-1 by \citet{cer21zp} with abundances of 1.4 $\times$ 10$^{-10}$ and 2.0 $\times$ 10$^{-10}$, respectively. We do not differentiate between these two forms in the network. 
In addition to the reaction between \ce{C2H} and \ce{c-C5H6} mentioned above, we have also included formation via the dissociative recombination of the \ce{C7H7+} ion which forms primarily from the reaction of \ce{CH2CCH+} with benzene \citep{smy82} as well as in ion-neutral reactions between \ce{c-C6H5+} and various hydrocarbons \citep{aus89}.

The detection of 1-cyanocyclopentadiene, \ce{C5H5CN}, in TMC-1 by \citet{mcc21} was followed quickly by the identification of 2-cyanocyclopentadiene \citep{lee21z}.  Lee et al. analysed their velocity-stacked spectra and used MCMC modelling to derive total fractional abundances of 8.3 $\times$ 10$^{-11}$ and 1.9 $\times$ 10$^{-11}$, respectively. 
We ignore the difference in structure between these two isomers in the database. Lee et al. suggest that the molecule is formed by the reaction of \ce{CN} with cyclopentadiene, c-C$_5$H$_6$, in analogy with the formation of benzonitrile from benzene, while \citet{cer21zp} suggest formation via the reaction between \ce{C2N} and \ce{CH2CHCHCH2}. 
Here we take the rate coefficient for the former reaction to be the same as that for benzonitrile formation. Given their large dipole moments \citep{lee21z}, proton transfer reactions and other ion-neutral rate coefficients are calculated using Eqn.~\ref{eqn:id}. 

\citet{cer21zp} tentatively detected ethynyl benzene, c-C$_6$H$_5$CCH, in TMC-1 at an abundance 2.5 $\times$ 10$^{-10}$. They suggested two formation routes, reactions of \ce{C2H} with \ce{C6H6} and \ce{C6H5CN}, where the first of these has been measured to have a fast reaction down to 103~K \citep{gou06}. Although not applicable in the case of dark clouds, we have added the measured high-temperature formation reaction between \ce{C2H2} and \ce{C6H5} to the network and adopted temperature-independent rate coefficients for its destruction.

Benzonitrile, \ce{C6H5CN} was first detected in TMC-1 by \citet{mcg18}. Subsequently, \citet{bur21za} detected it in four other cold sources and derived a total abundance of 1.6 $\times$ 10$^{-10}$ in TMC-1, using an MCMC analysis to fit their data to four components in their line-of-sight. 
Burkhardt et al. present a model for its interstellar synthesis. It is readily formed in the reaction of \ce{CN} with benzene, \ce{C6H6}, a reaction that is measured to be fast down to 15~K \citep{coo20}. Both its proton affinity and dipole moment are relatively large and we therefore allow proton-transfer reactions with \ce{H3+}, \ce{N2H+}, \ce{HCO+}, and \ce{H3O+} to produce \ce{C6H5CNH+} with rate coefficients calculated using Eqn.~\ref{eqn:id}. We note that proton transfer reactions, and therefore \ce{C6H5CNH+}, are not included in the Burkhardt et al. chemical model. We assume that its DR leads to benzonitrile and phenyl, \ce{C6H5}, with equal branching ratios.

Indene, c-C$_9$H$_8$, was detected in TMC-1 by both \citet{cer21zb} and \citet{bur21}. The latter authors have described a pathway to indene based on the work of \citet{dod21} and involving the reaction of \ce{CH} with styrene, c-C$_6$H$_5$C$_2$H$_3$. Their calculated abundances fall several orders of magnitude below that observed. We have not included indene in the database.

\citet{sit22} have recently searched for the isomers of cyanoindene, \ce{C9H7CN} in TMC-1, detecting 2-cyanoindene at a fractional abundance of 2.1 $\times$ 10$^{-11}$.  The formation of this molecule likely occurs in a manner similar to that of benzonitrile, that is, through a fast reaction between \ce{CN} and indene. Such a reaction results in an abundance ratio of cyanoindene to indene close to that observed but gas-phase synthesis, as currently understood, is not able to reproduce the observed abundance by several orders of magnitude. Cyanoindene is not included here.

The model failure to reproduce the large observed abundances of the benzene-related cyclic hydrocarbons is ultimately due to a lack of efficient, barrierless pathways to \ce{C6H6}, mainly due to the fact that the fast pathway to propene formation is no longer feasible as discussed in Sect.~\ref{sec:ch3chchcn} \citep{lin13}.  The identification of such pathways, or alternative routes to benzene, is a key issue for astrochemistry research.

\subsection{H$_2$CCHC$_3$N and HCCCHCHCN}
\label{sec:vca}

Vinylcyanoacetylene, \ce{H2CCHC3N}, and its isomer  cyanovinylacetylene, \ce{HCCCHCHCN}, the latter in its trans-(E) conformer, have been detected in TMC-1 by \citet{lee21zi}. They calculate abundances of 2 $\times$ 10$^{-11}$ for the former and 3 $\times$ 10$^{-11}$ for the latter as well as an upper limit of 2 $\times$ 10$^{-11}$ for the trans-(Z) conformer of cyanovinylacetylene. 
The neutral-neutral reactions that form these species are taken from Lee et al. to which we add our usual suite of ion-neutral and photoreactions.

\subsection{CH$_2$CHOH}
\label{sec:ch3cho}

Vinyl alcohol, \ce{CH2CHOH}, an isomer of acetaldehyde, \ce{CH3CHO}, has been detected in TMC-1 at an abundance of 2.5 $\times$ 10$^{-10}$ by \citet{agu21zj}, very close to the abundance, 3.5 $\times$ 10$^{-10}$, of acetaldehyde \citep{cer20d}. In general, experimental studies of neutral-neutral reactions cannot differentiate between isomeric products. 
As a result, it is possible that those NN reactions that are thought to produce \ce{CH3CHO} could also produce \ce{CH2CHOH}. The essential equality of the isomeric abundances suggests that both form by the same processes. {\sc Rate12} already contains several neutral-neutral reactions that form \ce{CH3CHO} so here we make the simple assumption that both isomers are produced by these reactions with equal branching ratios. We have made the same assumption for the DR of the protonated forms of both \ce{CH2CHOH} and \ce{CH3CHO}.

\subsection{HCCN, HCCO and HC$_4$N}
\label{sec:hccn}

Thirty years separate the discovery of \ce{HCCN} in the outer envelope of IRC+10216 \citep{gue91} and in TMC-1 \citep{cer21h}. Its synthesis has been discussed by \citet{osa04} and, with particular reference to Titan chemistry, by \citet{loi15}. Its formation is via neutral-neutral reactions:
\begin {eqnarray}
\mathrm{CH} + \mathrm{HCN} & \rightarrow & \mathrm{HCCN} + \mathrm{H} \\
\mathrm{CH} + \mathrm{HNC} & \rightarrow & \mathrm{HCCN} + \mathrm{H} \\
\mathrm{C} + \mathrm{H_2CN} & \rightarrow & \mathrm{HCCN} + \mathrm{H}, 
\end{eqnarray}
with loss through reactions with \ce{H}, \ce{N} and \ce{CH3} and with ions.

\ce{HCCO} was first detected in dark clouds by \citet{agu15b} and reported in TMC-1 by \citet{cer20d, cer21u} with a fractional abundance of 7.7 $\times$ 10$^{-11}$. Its chemistry was discussed by \citet{wak15b} who proposed both gas-phase and grain-surface syntheses. In the gas phase its major formation reactions are those of \ce{C2H} with \ce{OH} and \ce{O2}. For the former we adopt the rate coefficient suggested by Wakelam et al. 
For the latter, we include the four product channels in the compilation by \citet{bau05} with that to \ce{HCCO} given a branching ratio of 20\%.  Loss occurs through H and O atoms \citep{bau05} together with ion-neutral reactions. We follow the recommendation by \citet{wak15b} for the products and branching ratios for the dissociative recombination of \ce{H2CCO+}, replacing those contained in {\sc Rate12} which were taken from \citet{pra80}. \ce{HCCO} was not contained in {\sc Rate12} but its protonated form, \ce{CH2CO+}, was included.  For consistency, we have relabelled \ce{CH2CO+} as \ce{H2CCO+} in {\sc Rate22}.

\ce{HC4N} was detected in TMC-1 by \citet{cer21h} with an abundance of 3.7 $\times$ 10$^{-11}$.  We assume that it is formed in a similar fashion to \ce{HCCN}, that is through reactions of \ce{CH} and both \ce{HC3N} and its isomer \ce{HNC3}.
\begin {eqnarray}
\mathrm{CH} + \mathrm{HC_3N} & \rightarrow & \mathrm{HC_4N} + \mathrm{H} \\
\mathrm{CH} + \mathrm{HNC_3} & \rightarrow & \mathrm{HC_4N} + \mathrm{H} \\
\mathrm{CN} + \mathrm{C_3H_2} & \rightarrow & \mathrm{HC_4N} + \mathrm{H} \\
\mathrm{CN} + \mathrm{H_2CCC} & \rightarrow & \mathrm{HC_4N} + \mathrm{H} \\
\mathrm{C_2N} + \mathrm{C_2H_2} & \rightarrow & \mathrm{HC_4N} + \mathrm{H}, 
\end{eqnarray}
with rate coefficients taken from \citet{loi15}.  
We calculate loss rates in IN reactions with common ions and also include loss with N atoms.

\subsection{CH$_3$Cl}
\label{sec:ch3cl}

Methyl chloride was detected in the proto-binary source IRAS 16293-2422B by \citet{fay17}. The chemistry of this molecule and related species has been discussed in detail by \citet{ach17}  and we adopt their scheme for {\sc Rate22}.

\subsection{PH$_3$}
\label{sec:ph3}

Phosphine was not included in the {\sc Rate12} release as no route to \ce{PH4+} could be found. As the PH$_\mathrm{n}^+$ reactions with \ce{H2} are endothermic, formation of the P-H bond proceeds via the RA of \ce{P+} and \ce{PH+} with \ce{H2}. However, \ce{PH3+} does not react with \ce{H2} to form \ce{PH4+}. Here we circumvent this issue by incorporating MN reactions of \ce{PH3+} with anions to form \ce{PH3} directly. Its loss via ion-neutral reactions is taken from \citet{cha94} with its photodissociation rate from \citet{sil21}.

\subsection{ArH$^+$}
\label{sec:arh+}

\ce{ArH+} was detected by \citet{bar13} in the \object{Crab Nebula} and subsequently in absorption toward several Galactic molecular clouds by \citet{sch14}. Argon has an unusual interstellar chemistry in that its proton affinity is lower than that of \ce{H2} while its ionisation potential is larger than that of atomic hydrogen meaning that \ce{Ar} cannot be ionised by the interstellar UV field nor can \ce{ArH+} be formed through the proton transfer between \ce{Ar} and \ce{H3+}.  
Instead, \ce{ArH+} is formed by the reaction of \ce{Ar+} with \ce{H2} where the ion is the result of cosmic ray ionisation of \ce{Ar} and charge exchange between \ce{He+} and \ce{Ar} \citep{bab18}.   We have adopted the chemistry and rate coefficients proposed by Schilke et al. augmented by the laboratory measurements of the rate coefficients of \ce{ArH+} with \ce{H2} and \ce{CO} \citep{vil82}. Its DR rate coefficient has been calculated by \citet{abd18} who show that it is vanishingly small for vibrationally cold \ce{ArH+}  and we neglect it here. Schilke et al. show that that \ce{ArH+} is a tracer of gas with a low \ce{H2} fraction.

\subsection{Circumstellar chemistry}
\label{cse_chem}

In the following we describe the chemistries of elements that, due to their large depletion in interstellar clouds, are  predominantly found in molecular form in the the circumstellar environments.

\subsubsection{MgO and MgOH}
\label{sec:MgO}

MgO and MgOH are formed through NN reactions in either the hot inner winds of O-rich AGB stars or following grain disruption in post-shock gas, rate coefficients taken from the theoretical calculations by \citet{dec18a}. In general these reactions have large energy barriers. MgO also forms in the radiative association of Mg and O \citep{bai21a} but the rate coefficient is too small to produce it in any significant amount. Its photodissociation rate is taken from \citet{bai21b}.

\subsubsection{MgNC, MgCN and HMgNC}
\label{sec:mgnc}

These Mg-bearing species have all been detected in the outer envelope of the carbon-rich AGB star, IRC+10216 -- MgNC \citep{gue93}, MgCN \citep{ziu95}, and HMgNC \citep{cab13} -- indicating that a relatively low-temperature 
chemistry produces these species. 
Cabezas et al. calculate column densities  of 1.3 $\times$ 10$^{13}$ (\ce{MgNC}), 7.4 $\times$ 10$^{11}$ (\ce{MgCN}), and 6 $\times$ 10$^{11}$ (\ce{HMgNC}) cm$^{-2}$ to show their abundance ratios scale roughly as MgNC:MgCN:HMgNC = 15:1:0.8. 
The first two are likely formed in the dissociative recombination of \ce{MgNCH+} formed by the radiative association of \ce{Mg+} and \ce{HCN} \citep{pet00}. 
We used the Levenberg-Marquardt algorithm to fit their tabulated data over the temperature range 10--100~K by the expression 1.96 $\times$ 10$^{-17}$(\textrm{T}/300)$^{-1.49}$ cm$^3$ s$^{-1}$ and added the association of \ce{Mg+} with \ce{HNC} which is slightly faster than that with \ce{HCN}. 
Cabezas et al. also suggest that some \ce{HMgNC} could be formed with a branching ratio of about 1\% in the dissociative recombination of larger Mg-cyanopolyyne ions and we include this possibility (see Sect.~\ref{sec:mgcyano}). 
Mutual neutralisation between the protonated versions of these molecules with the observed anions, listed in Sect.~\ref{sec:intro}, are also included. For MN reactions involving \ce{MgNCH+} we have assumed a 50:50 branching ratio to \ce{MgNC} and \ce{MgCN}.

 We have included the standard set of IN reactions and ion-dipole rate coefficients for the loss of these species. For the case of IRC+10216, however, photodissociation is likely to be a more important loss process in the external envelope and one that can determine both column densities and the radial extent of these species. We have allocated uniform (and highly uncertain) rates of 10$^{-10}$\textrm{e}$^{-1.7{A_V}}$ \textrm{s}$^{-1}$ for all these species. 

\subsubsection{MgC$_3$N, HMgC$_3$N,  MgC$_5$N and MgC$_7$N}
\label{sec:mgcyano}

The magnesium cyanopolyynes, \ce{MgC3N} and \ce{MgC5N} have both been detected in the outer envelope of IRC+10216 with column densities of 9.3 $\times$ 10$^{12}$ and 4.7 $\times$ 10$^{12}$ cm$^{-2}$, respectively \citep{cer19, par21}. The formation of these species is thought to proceed from the radiative association of \ce{Mg+} with the HC$_{2\mathrm{n}+1}$N cyanopolyynes. Their rate coefficients have been tabulated by \citet{dun02} over the range 10-300~K. Our fits to these are shown in Table~\ref{tab:racyano}. For completeness, we include the rate coefficient for \ce{HC9N} in this table although \ce{MgC9N} is not included in the ratefile.  On the other hand, \ce{MgC7N}, which is not yet detected, is included here.

The large ions created via radiative association will dissociatively recombine with electrons and could, in principle, have several product channels. Here, we have followed the scaling procedures adopted by \citet{par21} with a total rate coefficient of 3 $\times$ 10$^{-7}$ (\textrm{T}/300)$^{-1/2}$ cm$^3$ s$^{-1}$. As an example, the dissociative recombination of \ce{MgC5NH+} leads to \ce{MgNC}, \ce{MgC5N}, and \ce{HMgNC} with values of the rate coefficients at 300~K of 7.80 $\times$ 10$^{-8}$, 2.22 $\times$ 10$^{-7}$, and 2.22 $\times$ 10$^{-9}$ cm$^3$ s$^{-1}$, respectively. 
In addition, we have included MN reactions between the observed anions and Mg-bearing cations. The Mg-carbon chain molecules have very large dipole moments, in excess of 6~D, and therefore very large rate coefficients for ion-neutral reactions at low temperature. As for the smaller Mg-CN species, their abundance and extent in the outer envelope is likely to be determined by photodissociation; we adopt the same photodissociation rates as those for the smaller Mg-bearing species. In all cases we assume that the product is \ce{Mg} and the appropriate C$_{2\textrm{n}+1}$N radical.

Recently \citet{cab23ae} detected \ce{HMgC3N} in IRC+10216 with a column density of 3 $\times$ 10$^{12}$ cm$^{-2}$. Its formation is uncertain but may occur as a product channel in the DR of the larger Mg-cyanopolyyne ions.  It is not included in {\sc Rate22}.

\begin{table}
 \caption{Fit parameters for the radiative association of \ce{Mg+} with cyanopolyynes over the range 10--300~K. The fits for HCN and HNC are over the range 10--100~K.}  
 \label{tab:racyano}
\begin{tabular}{llll}
\hline \hline
Neutral & $\alpha$ & $\beta$ & $\gamma$ \\
\hline
\ce{HCN} & 1.96 $\times$ 10$^{-17}$  & -1.49 & 0.0 \\
\ce{HNC} & 3.71 $\times$ 10$^{-17}$  & -1.48 & 0.0 \\
\ce{HC3N} & 7.75 $\times$ 10$^{-14}$ & -1.94 & 10.79 \\
\ce{HC5N} & 1.90 $\times$ 10$^{-10}$ & -2.24 & 33.6 \\
\ce{HC7N} & 3.98 $\times$ 10$^{-9}$ & -0.705 & 0.0 \\
\ce{HC9N} & 6.92 $\times$ 10$^{-9}$ & -0.48 & 0.0 \\
\hline
\end{tabular}
\end{table}

\subsubsection{MgC$_2$, MgC$_2$H, MgC$_4$H, MgC$_6$H and MgC$_8$H}
\label{sec:mgc2h}

The magnesium acetylide chains \ce{MgC2H}, \ce{MgC4H}, and \ce{MgC6H} have all been detected in the outer envelope of IRC+10216 \citep{agu14, cer19, par21} with column densities of 2 $\times$ 10$^{12}$, 2.2 $\times$ 10$^{13}$, and 2 $\times$ 10$^{13}$ cm$^{-2}$, respectively, and abundance ratios of 1:11:10 in order of increasing size \citep{par21}. 
More recently, \ce{MgC2} has been detected in the same region with a column density of 10$^{12}$ cm$^{-2}$ \citep{cha22}. Their formation routes are, as for the magnesium cyanopolyynes, thought to be dominated by radiative association, in this case between \ce{Mg+} and the polyacetylenes. 
These rate coefficients are tabulated by \citet{pet00} and \citet{dun02} and we have again found a best fit to them (Table~\ref{tab:raacet}). The ions so produced are assumed to recombine in a manner in which 80\% of the recombinations lose either a hydrogen or a magnesium 
atom, for example, dissociative recombination of \ce{MgC4H2+} leads to either \ce{Mg} + 
\ce{C4H2} or \ce{MgC4H} + \ce{H} with equal branching ratios, with the remaining 20\% divided equally between \ce{MgC2} and \ce{MgC2H}.  
Of these species, the most difficult to match calculated and observed column densities is \ce{MgC4H} for which the value of the RA rate coefficient is more than three orders of magnitude smaller than those for the larger hydrocarbons. 

In common with the magnesium cyanopolyynes, photodissociation plays an important role in setting the abundance and radial extent of these molecules. We assume the same rate coefficients as in Sect.~\ref{sec:mgnc} and assume that the sole product is \ce{Mg} and the appropriate C$_{2\textrm{n}}$H radical.

\begin{table}[h]
 \caption{Fit parameters for the radiative association of \ce{Mg+} with polyynes over the range 10--300~K. The fit for \ce{C2H2} is over the range 10--100~K.} 
 \label{tab:raacet}
\begin{tabular}{llll}
\hline \hline
Neutral & $\alpha$ & $\beta$ & $\gamma$ \\
\hline
\ce{C2H2} & 9.40 $\times$ 10$^{-18}$  & -1.47 & 0.0 \\
\ce{C4H2} & 2.57 $\times$ 10$^{-15}$ & -1.955 & 12.0 \\
\ce{C6H2} & 1.19 $\times$ 10$^{-12}$ & -2.21 & 21.21 \\
\ce{C8H2} & 3.41 $\times$ 10$^{-10}$ & -1.78 & 33.29 \\
\ce{C10H2} & 2.28 $\times$ 10$^{-9}$ & -0.60 & 10.48 \\
\hline
\end{tabular}
\end{table}

\subsubsection{Aluminium chemistry}
\label{sec:Al}

Seventeen Al-bearing molecules, including \ce{AlO}, \ce{AlOH}, \ce{AlF}, and \ce{AlCl} which are detected in AGB stars \citep{dec17, dan21}, are included in this release. Given the fact that along with Ti and Ca, aluminium is one of the most depleted elements in interstellar clouds, our primary purpose is to include a network that describes as best as possible the synthesis of observed Al-bearing molecules in hot gas in the inner circumstellar envelopes of these stars.  
As such, it is dominated by nearly 100 neutral-neutral reactions and contains molecules in size up to 5 atoms (\ce{Al2O3}) and in mass up to 132 amu (\ce{AlCl3}). Some of these, particularly the oxides, have measured rate coefficients but the majority are theoretical estimates which, because of the high-temperature environment in which these species are found, have both forward and backward determinations of rate coefficients.  
The main sources of information on these reactions comes are from \citet{swi03} and \citet{gob22}. We note that the latter authors, who are interested in how Al-O clusters can grow to dust grains, describe routes to molecules much larger than included in the database. 

There is no gas-phase formation for \ce{AlH3} included in {\sc Rate22}.  We retain it for those occasions in which it may be chosen as a parent molecule in the calculation of Al chemistry in the inner regions of O-rich AGB stars. If not included as a parent, its abundance is always calculated to be zero.

\subsubsection{Calcium chemistry}
\label{sec:Ca}

Eleven NN reactions are used to describe \ce{CaO} and \ce{CaOH} formation at high temperature, appropriate for conditions in the inner envelopes of O-rich AGB stars. Rate coefficients are a mix of theoretical, taken mainly from \citet{dec18a}, and experimental values.

\subsubsection{Titanium chemistry}
\label{sec:Ti}

\ce{TiO} and \ce{TiO2} are detected species in the circumstellar envelopes of O-rich AGB stars \citet{dec20} and \citet{wal23}. Laboratory data exists for the synthesis of both through a mixture of IN and NN chemistry.  
Rate coefficients for the latter are mostly experimental and taken from \citet{cam93} and \citet{pla13}. \ce{Ti} is reactive with several common interstellar and circumstellar oxides to form \ce{TiO} which can be converted to \ce{TiO2} through reaction with \ce{OH} and \ce{O2}. 
Some of these reactions are known to be efficient at room temperature and, since both have large dipole moments, we have included proton transfer with rate coefficients calculated using Eqn.~\ref{eqn:id} which, following DR, lead to loss mechanisms for both. A low-temperature interstellar synthesis of these species is therefore possible though this will be limited in effect since \ce{Ti} is one of the most heavily depleted elements in interstellar gas.

\subsubsection{Sodium chemistry}
\label{sec:nacn}

\citet{pet00} and \citet{dun02} give rate coefficients for the RA of \ce{Na+} with HCN and the cyanopolyynes up to \ce{HC9N} for temperatures up to 100~K and 300~K, respectively. As for the corresponding data for \ce{Mg+}, we have fitted their tabulated values to the usual modified Arrhenius formula.
\ce{NaCN} is a T-shaped molecule and has been observed in IRC+10216, originally by \citet{tur94} and mapped with ALMA by \citet{qui17} who show that, in common with \ce{CH3CN} \citep{agu15} and \ce{HC3N} \citep{sie22}, it has emission close to the star and extends out to a distance of a few times 10$^{15}$ cm. 

To estimate the abundance of \ce{NaCN}, which has a dipole moment of 8.85~D, we have included the usual set of ion-dipole rate coefficients and adopted an unshielded photodissociation rate of 10$^{-10}$ s$^{-1}$. The branching ratios of the DR reactions of the NaC$_{2\mathrm{n}+1}$NH$^{+}$ ions (n = 0--4) are unknown; here we adopt a 50:50 ratio between the \ce{Na} and \ce{NaCN} channels.

\citet{cab23ae} detected \ce{NaC3N} in the outer envelope of IRC+10216 with a column density of 1.2 $\times$ 10$^{11}$ cm$^{-2}$, about a factor of eight less than that of \ce{MgC3N} and a detection aided by its extremely large dipole moment, 12.9~D. Its formation routes are unknown but, as with the Mg cyanopolyynes, they are likely to involve the dissociative recombination of the products of the radiative association between \ce{Na+} and the cyanopolyynes. Table~\ref{tab:na_racyano} shows our fits to the rate coefficients calculated by \citet{dun02} for the radiative associations of \ce{Na+} with the cyanopolyynes.  Below 30~K, the rate coefficients for \ce{Na+} are, however, more than an order of magnitude smaller than those of \ce{Mg+} with \ce{HC3N} and \ce{HC5N} indicating that efficient formation of \ce{NaC3N} may depend on either a larger branching ratio or rate coefficent in the DR of \ce{NaC3NH+}, or on preferential production via DR of the larger RA products.  We note that \ce{NaC3N} is not yet included in the database.

Sodium chloride has been detected most frequently in the CSEs of both C-rich and O-rich AGB stars, \citep{cer87, dec16}, in the post-AGB object \object{CRL 2688} \citep{hig03} and in one early-type region, the protostellar disk around  \object{Orion SrcI} \citep{gin19}. 
The formation of  \ce{NaCl} is very poorly understood. \citet{sim22} have calculated the rate coefficient of the direct formation through the the RA between \ce{Na} and \ce{Cl} over the temperature range 30--750~K. 
We have also included formation via the reaction of \ce{Na} and \ce{HCl} for which we adopt a fit over 591-966~K to the two experimental data sets in the  NIST database.  \citet{pet96} has suggested that \ce{NaCl} can react with \ce{CN} to produce \ce{NaCN} and we have included this reaction with a (very uncertain) rate coefficient of 10$^{-10}$ cm$^3$ s$^{-1}$. \ce{NaCl} has a large dipole moment, 9.0~D, so ion-dipole rate coefficients are adopted, with photodissociation and cosmic ray-induced photodissociation rates taken from \citet{hea17}. 

Finally, it is worth pointing out that \ce{Na} has a very low ionization potential. However, \ce{Na+} is unreactive with many common neutral species which makes it an important source of free electrons in dark clouds.

\begin{table}
 \caption{Fit parameters for the radiative association of \ce{Na+} with cyanopolyynes over the range 10--300~K. The fits for HCN and HNC are over the range 10--100~K.}  
 \label{tab:na_racyano}
\begin{tabular}{llll}
\hline \hline
Neutral & $\alpha$ & $\beta$ & $\gamma$ \\
\hline
\ce{HCN} & 2.32 $\times$ 10$^{-18}$  & -1.50 & 0.0 \\
\ce{HNC} & 1.12 $\times$ 10$^{-17}$  & -1.50 & 0.0 \\
\ce{HC3N} & 3.19 $\times$ 10$^{-15}$ & -1.96 & 11.25 \\
\ce{HC5N} & 2.13 $\times$ 10$^{-12}$ & -2.65 & 30.09 \\
\ce{HC7N} & 3.18 $\times$ 10$^{-10}$ & -2.41 & 35.11 \\
\ce{HC9N} & 2.65 $\times$ 10$^{-9}$ & -1.24 & 17.18 \\
\hline
\end{tabular}
\end{table}

\subsection{Other detected molecules}

As noted previously, there are a number of molecules detected in recent years that are not included in {\sc Rate22}. The main reasons for such exclusion are: (i) that gas-phase formation routes have not yet been identified for them; (ii) their formation, while known, depends on the presence of another species whose origin is unknown; and (iii) grain surface chemistry likely dominates their syntheses.  
In addition to those molecules  mentioned previously, these include  ethylene oxide, also known as oxirane, \ce{c-C2H4O}, isocyanogen, \ce{CNCN}, detected in TMC-1 by \citet{agu18x}, nitrous acid, \ce{HONO}, 3-hydroxypropenal, \ce{HOCHCHCHO}, also known as malonaldehyde, detected in IRAS 16293-2422B \citep{cou19, cou22}, and methoxymethanol, \ce{CH3OCH2OH}, discovered in \object{NGC6334I MM1} \citep{mcg17b}. 
Missing species, ignoring conformers and the fullerene molecules, are shown in Table~\ref{tab:missing}.

\begin{table*}[th]
\centering
\caption{Detected species, listed by number of atoms, missing from the {\sc Rate22} database.  Tentative identifications are labeled by `?' \label{tab:missing}}
\begin{tabular}{lllllll}
\hline \hline
% Code & Reaction type & \multicolumn{2}{c}{Count} \\
\hline

 2/3 & \ce{SiP }& \ce{KCl} & \ce{SiCN} & \ce{SiCSi}  & \ce{AlNC}  & \ce{FeCN} \\
  & \ce{KCN} & \ce{CaNC} & \ce{HSC} & \ce{NCO} & \ce{HSO} \\
 4 & \ce{HOOH} & \ce{CNCN} & \ce{HNCN} & \ce{HONO} & \ce{NCCP}? &  \\
 5 & \ce{CNCHO} & \ce{HCOSH} & \ce{HNCNH} & \ce{NH2OH} & \ce{NH2CN} & \ce{NaC3N} \\
  & \ce{MgC3N+}  \\
 6 & \ce{CH3NC} & \ce{c-C3H2O} & \ce{CH2CNH} & \ce{SiH3CN} & \ce{NHCHCN} & \ce{NCCHNH} \\
  & \ce{MgC4H+}   & \ce{HMgC3N} & \ce{HOCOOH} & \ce{H2CCCN} \\
 7 & \ce{CH2OCH2} & \ce{HOCH2CN} & \ce{HCCCHNH} & \ce{c-C2H4O} & \ce{MgC5N+} &  \\
 8 & \ce{NH2CH2CN} & \ce{CH3CHNH} & \ce{CH3SiCH3} & \ce{NH2CONH2} & \ce{HCCCHCCC} & \ce{CH3CHCO}  \\
  & \ce{MgC6H+}  \\
 9 & \ce{CH3CONH2} & \ce{C2H5SH} & \ce{CH3NHCHO}? & \ce{HOCHCHCHO} &  &  \\
 10 & \ce{C2H5CHO} & \ce{(CH2OH)2} & \ce{CH3OCH2OH} & \ce{CH3CHCH2O} & \ce{CH2CCHC3N} & \ce{CH2CHCH2CN} \\
  & \ce{CH2C(CH3)CN} & \ce{C2H5NH2} & \ce{C2H5NCO} & \ce{C2H5CHO} & \ce{NH2C(O)CH2OH} \\
 11 & \ce{C2H5OCHO} & \ce{CH3COOCH3} & \ce{NH2CH2CH2OH} & \ce{C4H5CN} & \ce{CH3COCH2OH} &  \\
 >12 & \ce{C3H7CN} & \ce{C2H5OCH3} & \ce{c-C10H7CN} & \ce{c-C9H8} & \ce{c-C9H7CN} & \ce{C3H7OH}  \\

\hline
\end{tabular}
\end{table*}

\section{Model results}
\label{sec:models}

 In this section, we describe our model calculations for conditions appropriate to the dark cloud TMC-1 and to the circumstellar envelope of the carbon-rich AGB star IRC+10216. In both cases, we discuss the results for two ratefiles - our standard file and the reduced file in which the 53 endothermic reactions identified by \citet{tin23} are excluded (Sect.~\ref{sec:caveats}) and compare these to the extensive observational data for both objects.

\subsection{Dark cloud TMC-1}
\label{sec:tmc1_param}

We choose typical dark cloud parameters; n(H$_2$) = 10$^4$ cm$^{-3}$, a kinetic temperature T = 10~K, a visual extinction of 10 mag., and a cosmic-ray ionisation rate of 1.3 $\times$ 10$^{-17}$ s$^{-1}$.  

\subsubsection{O-rich TMC-1 model}
\label{sec:tmc1}

We have calculated a time-dependent model using the `low metal' initial elemental abundances given in Table~\ref{tab:tmc_initial}. These O-rich (C/O less than 1) abundances are the same as those in \citet{mce13} with the exception of S for which we adopt the value of 1.5 $\times$ 10$^{-6}$ as suggested by \citet{fue23} in their recent modelling of the chemistry of dark clouds in Taurus and Perseus.

\begin{table}[ht]
  \caption{Initial abundances relative to total H nuclei, n$_\mathrm{H}$.}
  \label{tab:tmc_initial}
  \begin{tabular}{l r l r} \hline \hline
    Species & $n_i/n_\mathrm{H}$\footnotemark[1] & Species $i$ & $n_i/n_\mathrm{H}$ \\ \hline 
H$_2$  &  0.5       &  Na  &  2.0(-09)  \\
H      &  5.0(-05)  &  Mg  &  7.0(-09)  \\
He     &  0.09      &  Si  &  8.0(-09)  \\
C      &  1.4(-04)  &  P   &  3.0(-09)  \\
N      &  7.5(-05)  &  S   &  1.5(-06)  \\
O      &  3.2(-04)  &  Cl  &  4.0(-09)  \\
F      &  2.0(-08)  &  Fe  &  3.0(-09)  \\
 \hline
  \end{tabular}
\tablefoot{
  \tablefoottext{1}{$a(b)=a \times 10^{b}$}
}
\end{table}

\begin{figure}
   \centering
   \includegraphics[width=\hsize]{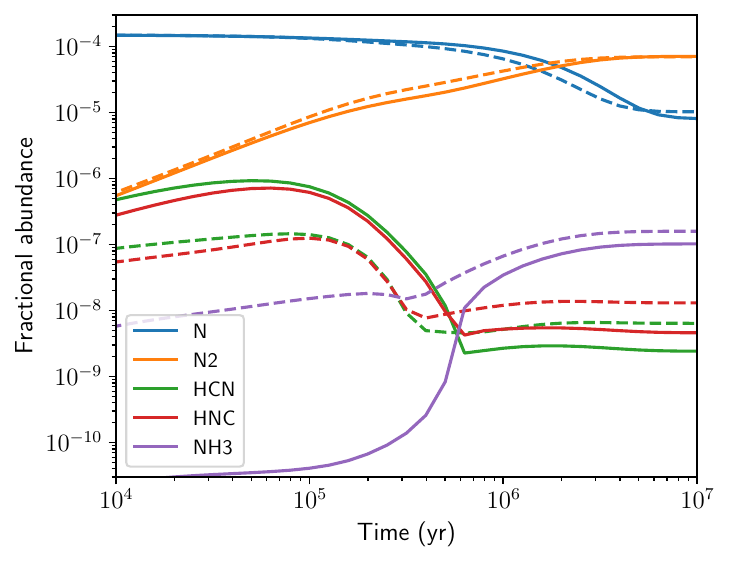}
      \caption{Time evolution of various N-bearing molecules in the dark cloud, O-rich model for both {\sc Rate22} (solid lines) and {\sc Rate12} (dashed lines).
             }
         \label{fig:NH3}
   \end{figure}

\begin{figure}
   \centering
   \includegraphics[width=\hsize]{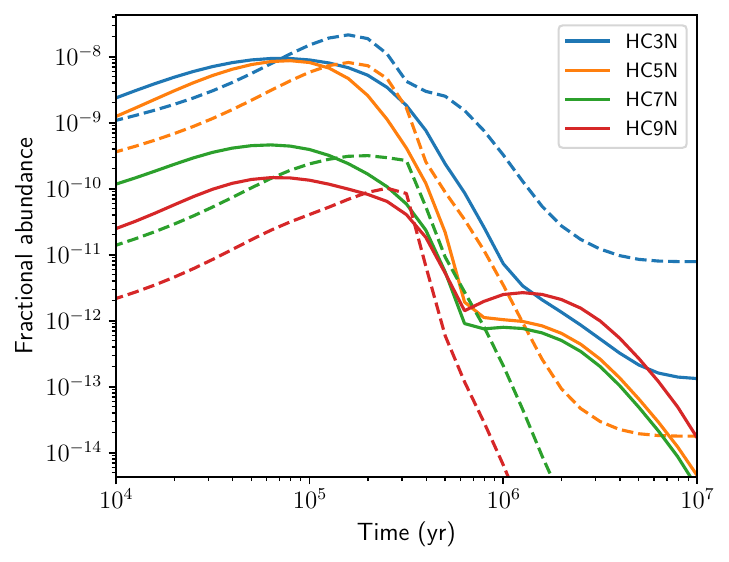}
      \caption{Time evolution of various cyanopolyynes in the dark cloud, O-rich model for both {\sc Rate22} (solid lines) and {\sc Rate12} (dashed lines).
              }
         \label{fig:HC3N}
   \end{figure}

Figures~\ref{fig:NH3} and \ref{fig:HC3N} show the time evolution of various N-bearing molecules in the O-rich, {\sc Rate22} and {\sc Rate12} models.  
There are significant differences between the results that are, in major part, driven by the inclusion of the reaction of \ce{C} + \ce{NH3} in {\sc Rate22}. This reaction was studied experimentally from 50~K to 300~K by \citet{hic15} and shown to be fast at low temperatures. 
Since atomic carbon is the dominant form of this element before it is processed into CO, the abundance of \ce{NH3} is depressed at times up to about 5 $\times$ 10$^5$ yrs.
For the cyanopolyynes, the situation is somewhat different. Here, their peak abundances in {\sc Rate22} are reached at slightly earlier times than in {\sc Rate12} but the peak abundances are fairly similar. The major differences now occur at long times, greater than 10$^6$ yrs.

\begin{figure}
   \centering
   \includegraphics[width=\hsize]{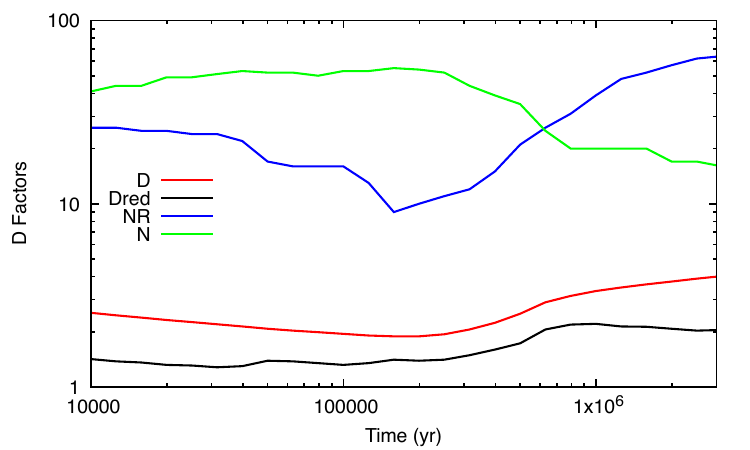}
      \caption{Best-fit, D, and reduced best-fit, Dred, average difference between observed and calculated species for the full ratefile and O-rich initial conditions. The curve labelled `NR' indicates the number of species removed in calculating the global average, Dred, and N is the number of species whose abundances lie within an order of magnitude of those observed. 
              }
         \label{fig:D}
   \end{figure}

We have searched for a `best time' fit between calculated, f$_c$, and observed, f$_o$, fractional abundances for those 134 species for which we have `retrieved'  abundances, that is, we neglect those species for which only upper limits are available.  To do so, we calculate the average modulus of the difference, $D$, in the logarithms of these between these two abundances:  

\begin{equation}
D = \frac{1}{N_{obs}} \Sigma ([\mathrm{log} f_{c} - \mathrm{log} f_{o}]^2)^{1/2}.
\end{equation}
Since we know that gas-phase chemistry is not appropriate for certain species - \ce{CH3OH} is a well-known example - we also calculated a `reduced' difference, $D_{red}$, in which we remove species with $D$ $>$ 4, that is, those whose calculated abundance differs by more than 4 orders of magnitude from that observed.  Fig.~\ref{fig:D} shows values of D, D$_{red}$ and the number of excluded species, NR, as a function of time in the chemical evolution. A `best-fit' solution would be one in which D is a minimum and N, the number of species whose abundances are within an order of magnitude of those observed, a maximum. This occurs in the range of (1-2) $\times$ 10$^5$ yrs with the formal `best fit' time at 1.6 $\times$ 10$^5$ yrs at which NR = 9 and 55 of the 134 species have calculated abundances that fall within an order of magnitude of their observed values (see Table~\ref{tab:fit}). A better fit, in the sense of a lower $D_{red}$, can be found at earlier time but at the expense of removing a significant number of species. If we seek to mimimise the number excluded, we again find that a time of 1.6 $\times$ 10$^5$ yrs is the preferred time. At times longer than 3 $\times$ 10$^5$ yrs, NR increases and N decreases so that at 10$^6$ yrs, D = 3.34, NR = 39 and N = 20, an extremely poor fit. This table also includes values in parentheses when the reduced reaction ratefile is used. One sees that the exclusion of reactions makes the fit slightly worse although the overall differences are marginal at 1.6 $\times$ 10$^5$ yrs. 

\begin{table}[ht]
  \caption{Fitting information on the comparison between observed and calculated abundances for the O-rich case.
   Values in parentheses are values when 53 reactions identified by  \citet{tin23} are set equal to zero. See text for further details.}
  \label{tab:fit}
  \begin{tabular}{lcccc} \hline \hline
    Time(yr) & D & D$_{red}$ & NR & N \\ \hline 
  6.3 $\times$ 10$^4$ & 2.03(2.14) & 1.38(1.44) & 16(18) & 52(45) \\
  1.6 $\times$ 10$^5$ & 1.89(1.95) & 1.41(1.46) & 9(10) & 55(53) \\
  5.0 $\times$ 10$^5$ & 2.51(2.52) & 1.73(1.72) & 21(22) & 35(35) \\
\hline
  \end{tabular}

\end{table}

Table~\ref{app:tmc_compar}  compares the model results for the full ratefile with the observed fractional abundances, including upper limits, of 149 species in TMC-1 at a time of 1.6 $\times$ 10$^5$ years. It is clear from this table that the model, which has an initial C/O abundance = 0.44, fails to reproduce the abundances of many species by several orders of magnitude, as indeed was the case for the {\sc Rate12} models.  These include species that are thought, from observations of massive star-forming regions, to be either formed on icy grains or from the evaporation products of such surface reactions, species such as methanol, methyl formate, and dimethyl ether.

In general, the C$_n$O abundances are well fit for n = 2, 3 but are severely over-produced for larger values as are the related species, HC$_n$O. This may be due to an over-estimate of the RA rate coefficients for the CO-hydrocarbon ion reactions since these are estimated from three-body experimental measurements \citep{ada89}. 

\begin{figure}
   \centering
   \includegraphics[width=\hsize]{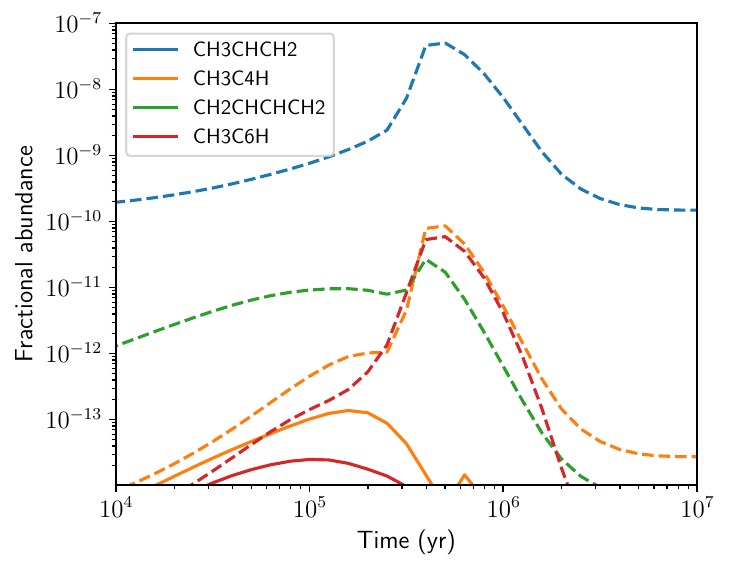}
      \caption{Time evolution of propene and species formed from it in the dark cloud, O-rich model for both {\sc Rate22} (solid lines) and {\sc Rate12} (dashed lines).
              }
         \label{fig:propene}
   \end{figure}

One of the  major changes between these results and {\sc Rate12} is the huge decrease in the calculated abundance of propene, \ce{CH3CHCH2}, by ten orders of magnitude, contrary to the fact that its observed value in TMC-1 is 4.0 $\times$ 10$^{-9}$ \citep{mar07}.   
Previously, it was formed by the DR of \ce{C3H7+} formed in successive radiative association reactions of \ce{H2} with \ce{C3H3+} and \ce{C3H5+}. These reactions, 
included in {\sc Rate12}, were subsequently determined to have barriers in their entrance channels and therefore not to proceed \citep{lin13}. 
The large decrease in the propene abundance has significant implications for larger hydrocarbons since it is a reactive species with large rate coefficients measured in NN reactions at low temperatures. For example, in terms of observed molecules in TMC-1, laboratory studies have shown that reactions of propene with \ce{C2} form \ce{CH3C4H} and its isomer allenyl acetylene, 
\ce{H2CCCHCCH}, those with \ce{C2H} form propyne, \ce{CH3CCH}, those with \ce{C4H} form \ce{CH3C6H}, those with \ce{CN} form vinyl cyanide, \ce{CH2CHCN}, crotononitrile, \ce{CH3CHCHCN}, methacrylonitrile, \ce{CH2C(CH3)CN}, and allyl cyanide, \ce{CH2CHCH2CN} \citep{cer22zo}, and those with \ce{CH} form cyclopentadiene, \ce{c-C5H6}, and butadiene, \ce{CH2CHCHCH2}. 
These newly formed species can then react to form other organic species. For example, \ce{c-C5H6} can react with CN to form cyanocyclopentadiene, \ce{C5H5CN}, with \ce{C2H} to form both fulvenallene, \ce{C5H4CCH2}, and ethynyl cyclopentadiene, \ce{C5H5CCH}, and with \ce{C} to form phenyl, \ce{c-C6H5} \citep{cer22zk}. 
Similarly, allenyl acetylene, \ce{H2CCCHCCH}, can react with \ce{CN} to form cyanoacetyleneallene, \ce{H2CCCHC3N}, an isomer of \ce{CH3C5N} \citep{shi21}. 
All of these species are calculated to have abundances at least three orders of magnitude less than their observed values in TMC-1. Very few of these large hydrocarbons are included in the {\sc Rate12} ratefile but, as an example of the different predictions of the {\sc Rate22} models, we compare in Figure~\ref{fig:propene} the abundances of four molecules in common, including those of propene and butadiene, \ce{CH2CHCHCH2}.  One sees clearly that this catastrophic decrease in the calculated propene abundance has a severe effect on those of the larger hydrocarbons.

\subsubsection{C-rich TMC-1 model}
\label{sec:crich}

\begin{figure}
   \centering
   \includegraphics[width=\hsize]{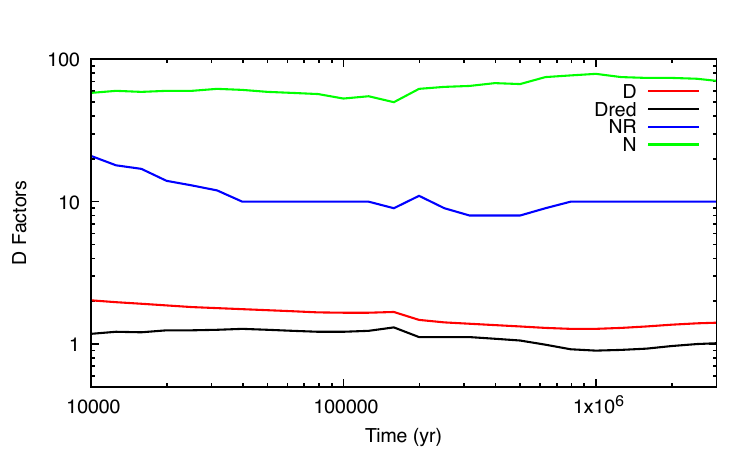}
      \caption{Best-fit, D, and reduced best-fit, Dred, average difference between observed and calculated species for the full ratefile and C/O = 1.4. The curve labelled `NR' indicates the number of species removed in calculating the global average, Dred, and N is the number of species whose abundances lie within an order of magnitude of those observed. 
              }
         \label{fig:D_Crich}
   \end{figure}

One possible mechanism to enhance the formation rate of complex hydrocarbons is to adopt a model with an elemental C/O ratio greater than 1, simulating an environment, say, where O is removed from the gas in the form of \ce{H2O} ice. We note that a selective depletion process for oxygen can also occur on warm dust in PDRs due to the high binding energy of O atoms \citep{esp19, rol22}.
Here we adopt the values for the fractional abundance of C and O relative to \ce{H2} of 2.8 $\times$ 10$^{-4}$ and 2.0 $\times$ 10$^{-4}$, respectively, values taken from \citet{agu13zg}.  
Fig.~\ref{fig:D_Crich} shows the results and Table~\ref{tab:fit_Crich} provides global information on the fit at various times.  
In general the D and D$_{red}$ values are reduced for the C-rich model compared to the O-rich model.  
For these conditions, the abundances of many of the complex hydrocarbons discussed above now agree with observation to within an order of magnitude at 10$^6$ yrs. In addition to propene, only \ce{CH3C4H}, \ce{c-C5H6}, \ce{CH3CHCHCN}, and \ce{C5H5CN} show differences with observation of more than an order of magnitude.

\begin{table}[ht]
  \caption{Fitting information on the comparison between observed and calculated abundances for the C-rich model with C/O = 1.4. Values in parentheses are for the case in  which C/O = 1.1. See text for further details.}
  \label{tab:fit_Crich}
  \begin{tabular}{lcccc} \hline \hline
    Time(yr) & D & D$_{red}$ & NR & N \\ \hline 
  6.3 $\times$ 10$^4$ & 1.70(1.75) & 1.24(1.28) & 10(10) & 58(60) \\
  1.6 $\times$ 10$^5$ & 1.68(1.67) & 1.31(1.23) & 9(10) & 50(52) \\
  5.0 $\times$ 10$^5$ & 1.33(1.22) & 1.06(0.90) & 8(9) & 67(80) \\
  1.0 $\times$ 10$^6$ & 1.28(1.28) & 0.90(0.90) & 10(10) & 79(80) \\
\hline
  \end{tabular}

\end{table}

In this case the number of species whose abundances agree to within an order of magnitude of those observed ranges from 50 at 1.6 $\times$ 10$^{5}$ yrs to a maximum of 79 at 10$^{6}$ yrs. In the latter case, only 10 species are different by more than four orders of magnitude and when these are removed, the average difference between observed and calculated values is less than a factor of 10.  
If one considers the additional 15 upper limits in Table~\ref{app:tmc_compar_Crich}, a further 7 species agree, that is the model gives agreement for 86 out of a total of 149 species, or 58\%. Many species showing the largest differences are the same in both models but in the C-rich case, there is a tendency to over-produce the large hydrocarbon chains, including \ce{C5H2}, \ce{C6H2}, C$_n$O (n = 4-7), HC$_n$O (n = 4-7) C$_n$S (n = 4,5), and the cyanopolyynes with n $>$ 7. 

Other values of C/O are used in the literature and to see the effect of these, we have also considered a model in which C/O = 1.1, with an O abundance of 3.1 $\times$ 10$^{-4}$ \citep{bar22}.  The models show an improvement with respect to the C/O = 1.4 models (see Table~\ref{tab:fit_Crich}) and in particular at a time of 5.0 $\times$ 10$^{5}$ yr, when D$_{red}$ is about 15\% smaller and N about 19\% larger than the values for C/O = 1.4.  In part, the improved fit is a result of a reduction in the over-production of large hydrocarbon chains discussed in the previous paragraph. 
As was also found for the O-rich case, the removal of the 53 endothermic reactions makes only a slight change in the results presented here for the full ratefile.

\subsection{IRC+10216}
\label{sec:cwleo}

We have also investigated the application of the ratefile to a model of the circumstellar envelope surrounding IRC+10216 in which the external interstellar UV field drives the chemistry.  We assume standard physical parameters for the outflow. Motivated by the revised, and larger, distance to IRC+10216 suggested by \citet{and22}, we adopt a uniform mass-loss rate, $\dot{M}$,  of 3.0 $\times$ 10$^{-5}$ M$_{\odot}$ yr$^{-1}$ (Table~\ref{tab:cwleo_hi_mlr}).  
We choose an expansion velocity, $v_e$, of 14.5 km s$^{-1}$, and a power-law temperature distribution, T(r) = T$_*$(r/R$_*$)$^{-0.7}$, with  a stellar temperature T$_*$ = 2330~K and stellar radius R$_*$ = 5.0 $\times$ 10$^{13}$ cm.  With a constant mass-loss rate and expansion velocity, the number density of \ce{H2}, which is proportional to $\dot{M}/v_e$, follows an $r^{-2}$ power law while the visual extinction to interstellar UV radiation follows an $r^{-1}$ distribution.  We have adopted input parent abundances from \citet{agu20} and \citet{vds21} augmented by metal atom abundances (Table~\ref{tab:cwleo_initial}). We calculate the chemistry from an initial radius of 10$^{14}$ cm out to 10$^{18}$ cm. At 10$^{14}$ cm, n(\ce{H2}) = 2.33 $\times 10^9$ cm$^{-3}$, T = 1424~K ,and A$_V$ = 1160 mag.

\begin{table*}[ht]
  \begin{center}
  \caption{Initial fractional abundances, $f$, relative to H$_2$ for the CSE model. $a(b)=a \times 10^{b}$.}
  \label{tab:cwleo_initial}
  \begin{tabular}{l l l l l l} \hline \hline
    Species & f & Species & f & Species & f  \\ 
    \hline 
He  & 1.70(-01)  &  CS & 1.06(-05) & SiO & 5.02(-06)  \\
SiS  &  5.98(-05) & CO & 8.00(-04) & \ce{C2H2} & 4.38(-05) \\
HCN & 4.09(-05) & \ce{N2} & 4.00(-05) & \ce{CH4} & 3.50(-06) \\
\ce{NH3} & 6.00(-08) & \ce{H2O} & 2.55(-06) & \ce{SiC2} & 1.87(-05) \\
\ce{HCl} & 3.25(-07) & HCP & 2.50(-08) & \ce{C2H4} & 6.85(-08) \\
HF & 1.70(-08) & \ce{H2S} & 4.00(-09) & \ce{SiH4} & 2.20(-07) \\
\ce{PH3} & 1.00(-08) & Na & 1.00(-09) & Mg & 1.00(-05) \\
Al & 1.00(-07) & Ar & 1.00(-08) & Ca & 1.00(-09) \\
Ti & 1.00(-09) & Fe & 6.00(-09) &  \\
 \hline
  \end{tabular}
  \end{center}
\end{table*}

 \begin{figure}
   \centering
   \includegraphics[width=\hsize]{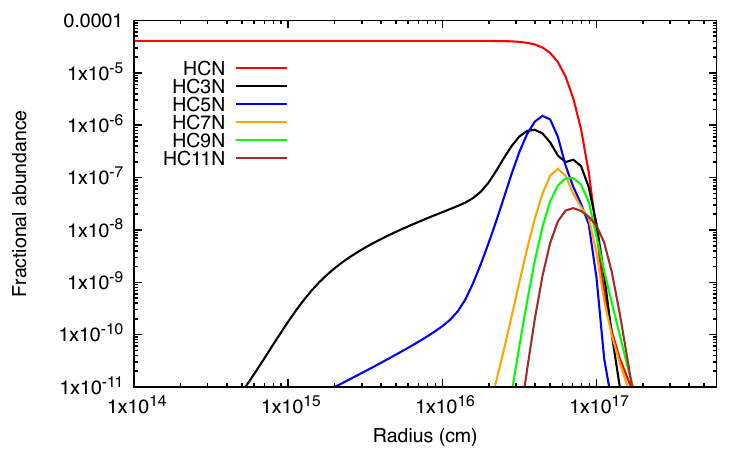}
      \caption{Fractional abundances of the cyanopolyynes for a mass-loss rate of 3.0 $\times$ 10$^{-5}$ M$_{\odot}$ yr$^{-1}$ as a function of radial distance from the central star. }
         \label{fig:cyano}
   \end{figure}

 \begin{figure}
   \centering
   \includegraphics[width=\hsize]{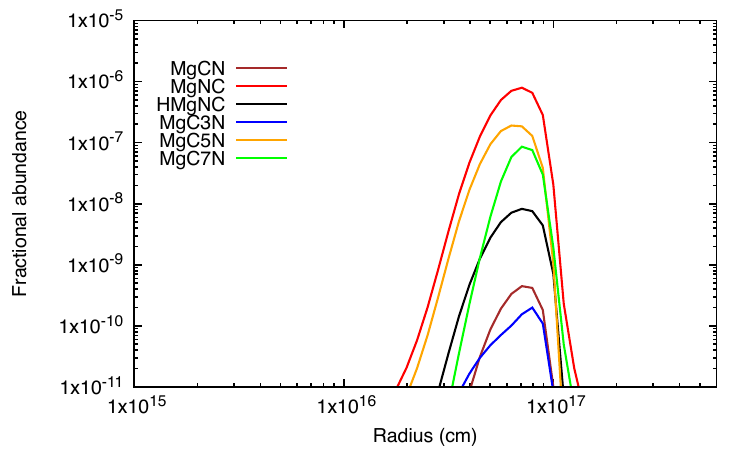}
      \caption{Fractional abundances of the Mg-terminated cyanopolyynes for a mass-loss rate of 3.0 $\times$ 10$^{-5}$ M$_{\odot}$ yr$^{-1}$ as a function of radial distance from the central star.}
         \label{fig:mgcyano}
   \end{figure}

Figures~\ref{fig:cyano} and \ref{fig:mgcyano} show the fractional abundances of the cyanopolyynes and the Mg-terminated cyanopolyynes as a function of radial distance.
We compare calculated column densities with those of 61 observed daughter species, that is, species which are not parents, in IRC+10216 for a mass-loss rate of 3.0 $\times$ 10$^{-5}$ M$_{\odot}$ yr$^{-1}$ in Table~\ref{tab:cwleo_hi_mlr}.  
We note, however, that observationally-derived column densities of the same molecule can differ by more than an order of magnitude as they are often dependent on a variety of factors: the particular excitation properties associated with observed transitions; the angular resolution of the telescope; whether or not the column density calculation includes an estimation of the size of the emitting region, often taken to be the photodissociation radius; and whether or not a beam-averaged column density is given. Thus our comparison is at best only indicative of how the model behaves.

We find that only one species, \ce{MgC4H}, has a column density more than 4 orders of magnitude different (less) than that observed.  
For our power-law temperature distribution, we find $D$ = 1.00(0.99), $D_{red}$ = 0.95(0.94), and that 37(38) of the 61 species have calculated and observed column densities that differ by less than an order of magnitude, where the values in parentheses are from the reduced ratefile. One sees that the reduction in the number of reactions has a negligible effect on our overall fit, with negligible differences also on column densities (though not shown here).

 Eight species have column densities that fall more than a factor of 100 less than those observed: the anions \ce{CN-}, \ce{C4H-}, and \ce{C7N-}; the Mg-species \ce{MgC4H} and \ce{MgC3N}, \ce{CH2NH}, \ce{CH2CN}, and \ce{CCP}. No species has a column density more than 100 times larger than observed.

\begin{figure}
   \centering
   \includegraphics[width=\hsize]{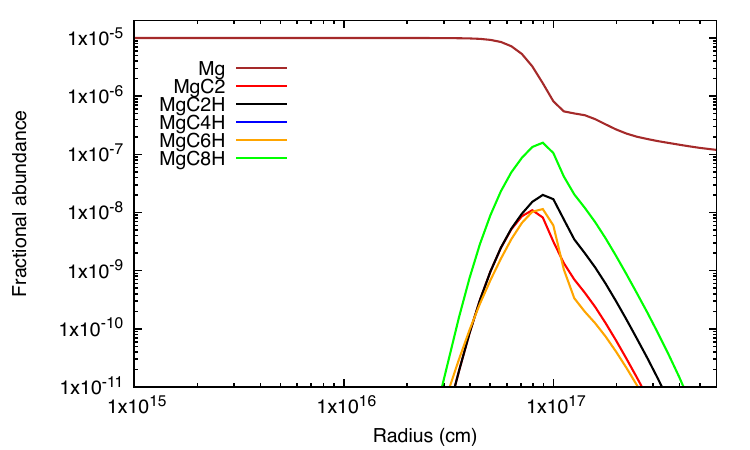}
      \caption{Fractional abundances of the Mg-terminated polyacetylides for a mass-loss rate of 3.0 $\times$ 10$^{-5}$ M$_{\odot}$ yr$^{-1}$ as a function of radial distance from the central star. }
         \label{fig:mg_hydrides}
   \end{figure}

Fig.~\ref{fig:mg_hydrides} shows the radial distribution of the magnesium polyacetylides for a mass-loss rate of 3.0 $\times$ 10$^{-5}$ M$_{\odot}$ yr$^{-1}$. The abundance of the longest chain \ce{MgC8H} is larger than the others primarily because the rate coefficient of the RA between \ce{Mg+} and \ce{C8H2} is larger than those of \ce{Mg+} with the smaller cumulenes (see Sect.~\ref{sec:mgc2h}).

The reason for the large discrepancy between model and observed abundances for \ce{MgC4H} could be due to the fact that we do not differentiate between diacetylene, \ce{HC4H}, and the cumulene \ce{H2C4}.  
As a result, we do not include either of these as a parent molecule whereas in fact \citet{fon18} have found that diacetylene has a column density of around 1.9 $\times$ 10$^{16}$ cm$^{-2}$ close to the star.  We thus investigated a model in which we included diacetylene as a parent species with an initial abundance of 1.6 $\times$ 10$^{-6}$ but found that the column density of \ce{MgC4H} increased only by an order of magnitude, still far too low. The reason appears to be that the RA rate coefficient (see Table~\ref{tab:raacet}) is simply too low.  
It also may be possible to synthesise the correct abundance if the larger protonated Mg-acetylide chains formed it via DR. That is, \ce{MgC6H2+}/\ce{MgC8H2+} + \ce{e-} $\rightarrow$ \ce{MgC4H} as well as other products. This would imply substantial rearrangement in the DR process and should be the subject of theoretical or experimental investigation.

\begin{table*}[t]
  \begin{center}
  \caption{Observed and calculated column densities (cm$^{-2}$) in IRC+10216 for a mass-loss rate of 3.0 $\times$ 10$^{-5}$ M$_{\odot}$ yr$^{-1}$. The `Ref.' column gives references for the observed values. $a(b)=a \times 10^{b}$. }

  \label{tab:cwleo_hi_mlr}
  \begin{tabular}{lrrclrrc} 
\hline \hline
Species   & Observed   &  Calculated  & Ref.   & Species  & Observed  & Calculated  &Ref.  \\
    \hline 
\ce{C} & 1.10(16) & 1.50(16) & A & \ce{C2} & 7.90(14) & 4.95(15) & B \\ 
\ce{C2H} & 4.60(14) & 3.36(15) & ZE & \ce{CN-} & 5.00(12) & 2.67(10) & C \\ 
\ce{CN} & 1.10(15) & 5.86(15) & B & \ce{HNC} & 1.10(14) & 2.77(14) & ZE \\ 
\ce{CH2NH} & 9.00(12) & 4.10(10) & T & \ce{HCO+} & 3.50(12) & 5.49(11) & A, V \\ 
\ce{H2CO} & 5.00(12) & 5.71(11) & U & \ce{C3} & 1.00(15) & 2.64(14) & A \\ 
\ce{C3H} & 5.60(15) & 2.00(14) & ZE & \ce{C3H2} & 2.00(13) & 5.54(13) & A \\ 
\ce{H2CCC} & 3.00(12) & 1.68(13) & W & \ce{HCCN} & 1.40(12) & 1.89(13) & ZD \\
\ce{SiC} & 6.00(13) & 2.31(14) & N & \ce{CH3CCH} & 1.80(13) & 8.85(11) & E \\ 
\ce{CH2CN} & 8.40(12) & 6.54(09) & A, E & \ce{CH3CN} & 4.30(13) & 8.66(12) & ZE \\
\ce{SiN} & 3.80(13) & 3.25(12) & Q & \ce{CP} & 1.00(14) & 2.27(12) & L \\ 
\ce{HCP} & 1.00(15) & 5.86(15) & L & \ce{PN} & 1.00(13) & 2.03(12) & L \\ 
\ce{H2CS} & 1.00(13) & 1.01(12) & E & \ce{MgC2} & 1.00(12) & 1.51(12) & ZB \\ 
\ce{C4H} & 5.50(15) & 6.17(14) & A & \ce{MgC2H} & 2.00(12) & 2.69(12) & Y \\ 
\ce{C4H-} & 7.00(11) & 3.31(08) & F & \ce{C3N} & 3.10(14) & 8.55(14) & H \\ 
\ce{C3N-} & 1.60(12) & 1.87(11) & H & \ce{MgCN} & 7.40(11) & 6.79(10) & ZA \\ 
\ce{MgNC} & 1.30(13) & 1.43(14) & ZA & \ce{HMgNC} & 6.00(11) & 1.54(12) & ZA \\ 
\ce{C4H2} & 1.60(14) & 6.38(14) & X & \ce{HC3N} & 1.50(15) & 4.71(14) & A \\ 
\ce{C3O} & 1.00(12) & 5.80(10) & S & \ce{CH2CHCN} & 5.00(12) & 1.58(12) & E \\ 
\ce{SiNC} & 2.00(12) & 3.81(10) & R & \ce{CCP} & 1.20(12) & 9.28(09) & M \\
\ce{C2S} & 1.50(14) & 7.81(13) & I & \ce{C5} & 1.00(14) & 2.30(12) & A \\ 
\ce{C5H} & 2.60(14) & 8.55(12) & A & \ce{HC4N} & 1.50(12) & 1.53(13) & ZD \\
\ce{SiC3} & 4.30(12) & 1.06(14) & O & \ce{C3S} & 8.50(13) & 7.96(13) & I, J \\ 
\ce{MgC4H} & 2.20(13) & 1.37(09) & Y & \ce{C6H-} & 4.00(12) & 4.22(11) & E \\ 
\ce{C6H} & 7.00(13) & 1.03(13) & E & \ce{C5N} & 4.50(12) & 5.61(13) & A, D \\ 
\ce{C5N-} & 3.00(12) & 2.51(12) & G & \ce{MgC3N} & 9.30(11) & 3.07(10) & Y \\ 
\ce{HC5N} & 2.50(14) & 3.76(14) & A & \ce{SiC4} & 7.00(12) & 2.17(13) & P \\ 
\ce{C7H} & 1.50(12) & 1.05(12) & A, D & \ce{C5S} & 3.60(13) & 1.28(14) & J \\ 
\ce{C8H-} & 2.00(12) & 1.90(11) & K & \ce{C8H} & 5.00(12) & 3.99(12) & A \\ 
\ce{MgC6H} & 2.00(13) & 1.44(12) & Z & \ce{MgC5N} & 4.70(12) & 3.76(13) & Z \\ 
\ce{C7N-} & 2.40(12) & 1.67(10) & ZC & \ce{HC7N} & 1.00(14) & 2.54(13) & A \\ 
\ce{HC9N} & 3.00(13) & 1.78(13) & A &   &   &   & \\

\hline
  \end{tabular}

   \end{center}

References: A: See references in Table 5 of \citet{mil00}: B: \citet{bak97}; C: \citet{agu10b}; D: \citet{cer00}; E: \citet{agu08} F: \citet{cer07}; G: \citet{cer08}; H: \citet{tha08}; I: \citet{cer87b}; J: \citet{bel93};  K: \citet{rem07zh}; L: \citet{mil08}; M: \citet{hal08}; N: \citet{cer89}; O: \citet{app99}; P: \citet{ohi89}; Q: \citet{tur92}; R: \citet{gue04}; S: \citet{ten06}; T: \citet{ten10}; U: \citet{for04}; V: \citet{pul11}; W: \citet{cer91a}; X: \citet{cer91b}; Y: \citet{cer19}; Z: \citet{par21}; ZA: \citet{cab13}; ZB: \citet{cha22}; ZC: \citet{cer23zt}; ZD: \citet{cer04};  ZE: Tuo et al., in prep. (2023).

\end{table*}
%

%-------------------------------------------------------------
%                              Table longer than a single page  
%-------------------------------------------------------------
% All long tables will be placed automatically at the end of the document
%

\section{Summary}
\label{sec:summary}

We have outlined the major differences between our new release of the UMIST Database for Astrochemistry, {\sc Rate22}, and the previous version {\sc Rate12}. The numbers of species and reactions have increased substantially in large part driven by the substantial number of new detections made over the past decade. Many observed species are, however, excluded from the database due to the lack of identified gas-phase mechanisms. 

We have calculated time-dependent chemical kinetic models for a dark cloud with physical conditions similar to those in TMC-1 for both O-rich and C-rich initial conditions.  As expected, the C-rich models give a much better agreement with observations with 70-80 having abundances within an order of magnitude of the 134 observed.  Our results, at least in terms of our `reduced least squares' fit, are not affected significantly by the choice of C/O ratio although the lower value of 1.1 reduces the degree of over-production of hydrocarbon chain molecules. 
We discuss reasons for the largest discrepancies and conclude that while grain surface formation is indicated for some species, the inefficient production of \ce{C3H7+} and hence of propene, \ce{CH3CHCH2}, is a key inhibitor of the production of larger hydrocarbons.  We have also applied our network to the case of an outflow from a carbon-rich AGB star, finding agreement to within an order of magnitude between calculated and observed column densities for about 60\% of the daughter species included in chemistry.  

In several cases, abundances are either greatly over- or under-produced. For dark clouds, these include species affected by the `propene catastrophe' as discussed in Sect.~\ref{sec:tmc1}. Other examples include the C$_n$O and the HC$_n$O species also discussed in Sect.~\ref{sec:tmc1}, and the Mg-bearing molecules observed in IRC+10216 (Sect.~\ref{sec:cwleo}). 
The chemistry of all these species rely on two processes, radiative association and dissociative recombination, and their calculated abundances are therefore very sensitive to the rate coefficients adopted for both and to branching ratios of the DR reactions.

We have also considered the removal of the endothermic reactions identified in the KIDA database by \citet{tin23}. We find that these make only a minor change to our global fits for all models considered here.

We have not attempted to produce isotopic versions of {\sc Rate22}, the most important of which would include $^{13}$C and D. Both would need significant effort if multiply-substituted species were to be included. Even if one considers single isotopic substitutions, there are severe uncertainties in the choice of additional reactions and rate coefficients. For example, during a formation process that involves reactants containing multiple carbon atoms, there is almost no information on where a $^{13}$C atom might replace a $^{12}$C atom, nor where $^{13}$C might be retained in the DR of a complex hydrocarbon ion.  
Nearly 200 species in {\sc Rate22} contain 5 or more C atoms. With the exception of \ce{HC5N} and \ce{HC7N}, \citep{bur18}, there is little observational data available on $^{13}$C/$^{12}$C ratios in larger molecules. Molecules with large numbers of carbon atoms are more likely to contain a $^{13}$C atom simply from a statistical viewpoint \citep{bur18}, further confusing an understanding of the impact of $^{13}$C chemistry.

For deuterium, the situation is complicated by the need to include the spin chemistry of \ce{H2}, its ions and deuterated counterparts, as well as those of multiply-deuterated species given that many such species are detected in interstellar clouds. \citet{maj17} published such a model based on a KIDA network that incorporated 7509 gas-phase reactions. The ratefile contained 111,000 reactions when deuterated. A similar multiplicative factor might be expected for the {\sc Rate22} ratefile.

In addition to the gas-phase chemistry discussed here, we have produced a website that will include the codes needed to generate chemical models for both interstellar clouds and circumstellar envelopes.  Previously, Van de Sande \& Millar extended the {\sc Rate12} CSE models to include the effects of porosity, clumpy outflows and irradiation by internal UV photons from the AGB star \citep{vds19} and from an nearby companion star \citep{vds22}. We shall extend the capability of our CSE models by calculating these photorates for use with the {\sc Rate22} network.
Finally, we note that, although there has been a decade between this and the previous release of the UDfA, we plan to issue updates to the database on an annual basis.

\begin{acknowledgements}
      We are extremely grateful to the anonymous reviewer whose close, careful reading and expert comments greatly improved the original manuscript. This work could not have been completed without the support of the Leverhulme Trust through the award of an Emeritus Fellowship to TJM.  He also acknowledges support from the STFC through grants ST/P000312/1 and ST/T000287/1. MVdS acknowledges support from the  European Union's Horizon 2020 research and innovation programme under the Marie Skłodowska-Curie grant agreement No 882991. CW acknowledges financial support from the University of Leeds, the Science and Technology Facilities Council, and UK Research and Innovation (grant numbers ST/T000287/1 and MR/T040726/1).
\end{acknowledgements}

% WARNING
%-------------------------------------------------------------------
% Please note that we have included the references to the file aa.dem in
% order to compile it, but we ask you to:
%
% - use BibTeX with the regular commands:
%   \bibliographystyle{aa} % style aa.bst
%   \bibliography{Yourfile} % your references Yourfile.bib
%
% - join the .bib files when you upload your source files
%-------------------------------------------------------------------

%
\bibliographystyle{aa}
\bibliography{rate22_revision}
%\bibliography{paper}
%

\begin{appendix}

\onecolumn

\section{Comparison of observed and calculated abundances in TMC-1}

We compare observed fractional abundances in TMC-1 with calculated abundances for two models, the first for the O-rich conditions described in Sect.~\ref{sec:tmc1} (Table~\ref{app:tmc_compar}), the second for the C-rich conditions described in Sect.~\ref{sec:crich} (Table~\ref{app:tmc_compar_Crich}).

%  \longtab[1]{
\begin{longtable}{lrrclrrc}
\caption{\label{app:tmc_compar} Observed fractional abundances, relative to H$_2$, in TMC-1(CP) and corresponding calculated `best fit' values at 1.6 $\times$ 10$^5$ years for the O-rich model. The `Ref.' column gives references for the observed values. $a(b)=a \times 10^{b}$. }\\

%\begin{tabular}{lrrclrrc}

\hline\hline
Species   & Observed   &  Calculated  & Ref.   & Species  & Observed  & Calculated  &Ref.  \\
\hline
\endfirsthead
\caption{continued.}\\
\hline\hline
Species   & Observed   &  Calculated  & Ref.   & Species  & Observed  & Calculated  &Ref.  \\
\hline
\endhead
\hline
\endfoot

\ce{OH} & 3.0(-07) & 1.9(-08) & ZG & \ce{NH3} & 5.0(-08) & 5.3(-11) & ZC \\ 
\ce{H2O} & <7.0(-08) & 2.1(-06) & ZG & \ce{C2H} & 6.5(-08) & 2.5(-09) & ZK \\ 
\ce{CN} & 5.0(-09) & 2.0(-08) & ZY & \ce{HCN} & 1.1(-08) & 4.3(-07) & ZW \\ 
\ce{HNC} & 2.6(-08) & 3.6(-07) & ZW & \ce{HCNH+} & 1.9(-09) & 2.5(-09) & ZG \\ 
\ce{H2CN} & 1.5(-11) & 1.8(-11) & ZG & \ce{H2NC} & <3.2(-11) & 9.0(-12) & ZM \\ 
\ce{CO} & 8.0(-05) & 1.7(-04) & ZW & \ce{HCO} & 1.1(-10) & 1.2(-09) & U \\ 
\ce{HCO+} & 9.3(-09) & 5.1(-09) & ZW & \ce{N2H+} & 4.0(-10) & 4.5(-11) & ZY \\ 
\ce{CH2NH} & <3.6(-09) & 1.5(-11) & ZU & \ce{NO} & 3.0(-08) & 2.3(-09) & ZY \\ 
\ce{H2CO} & 5.0(-08) & 3.7(-08) & F & \ce{H2COH+} & <3.0(-11) & 3.1(-11) & D \\ 
\ce{CH3OH} & 4.8(-09) & 1.7(-15) & D & \ce{H2S} & 5.0(-10) & 3.5(-13) & ZY \\ 
\ce{C3H} & 3.0(-09) & 8.5(-10) & ZC & \ce{C3H+} & 2.4(-12) & 7.9(-13) & G \\ 
\ce{C3H2} & 1.9(-09) & 1.4(-09) & ZC & \ce{H2CCC} & 1.9(-10) & 1.7(-11) & K \\ 
\ce{H2CCCH+} & 7.0(-11) & 1.3(-10) & YB & \ce{CH2CCH} & 1.0(-08) & 3.2(-11) & T \\ 
\ce{HCCN} & 4.4(-11) & 1.1(-09) & H & \ce{CH3CCH} & 1.2(-08) & 5.6(-12) & ZC \\ 
\ce{C2O} & 7.5(-11) & 8.9(-11) & U & \ce{CH2CN} & 1.5(-09) & 8.3(-13) & O \\ 
\ce{HCCO} & 7.7(-11) & 7.8(-14) & U & \ce{CH3CN} & 4.7(-10) & 2.8(-08) & O \\ 
\ce{CH2CO} & 1.4(-09) & 8.9(-09) & D & \ce{CH3CHCH2} & 4.0(-09) & 8.5(-22) & ZZ \\ 
\ce{HNCO} & 1.3(-09) & 4.5(-11) & D & \ce{HCNO} & 7.0(-12) & 2.1(-13) & D \\ 
\ce{HOCN} & 1.0(-11) & 5.4(-13) & D & \ce{CH3CO+} & 3.2(-11) & 1.6(-11) & R \\ 
\ce{CS} & 3.5(-08) & 6.6(-08) & B & \ce{CH3CHO} & 3.5(-10) & 2.1(-12) & D \\ 
\ce{CH2CHOH} & 2.5(-10) & 3.8(-12) & ZJ & \ce{H2NCO+} & <4.0(-12) & 6.1(-14) & D \\ 
\ce{HCS} & 5.5(-10) & 1.2(-11) & B & \ce{HCS+} & 1.0(-09) & 5.4(-11) & A \\ 
\ce{NH2CHO} & <5.0(-12) & 0 & D & \ce{HCO2+} & 4.0(-11) & 5.6(-12) & D \\ 
\ce{HCOOH} & 1.4(-10) & 6.1(-10) & D & \ce{CH3OCH3} & 2.5(-10) & 5.5(-19) & ZJ \\ 
\ce{C2H5OH} & 1.1(-10) & 4.0(-14) & ZV & \ce{H2CS} & 4.7(-09) & 4.3(-11) & B \\ 
\ce{NS} & 1.7(-10) & 1.7(-12) & E & \ce{NS+} & 5.2(-12) & 4.3(-12) & E \\ 
\ce{SO} & 1.0(-08) & 3.7(-10) & ZY & \ce{C4H} & 8.5(-09) & 2.6(-09) & YA \\ 
\ce{C4H-} & 2.1(-12) & 1.2(-12) & ZK & \ce{C3N} & 1.2(-09) & 1.6(-10) & YA \\ 
\ce{C3N-} & 6.4(-12) & 3.1(-13) & YA & \ce{C4H2} & 3.3(-10) & 2.3(-11) & K \\ 
\ce{HC3N} & 2.3(-08) & 6.8(-09) & C & \ce{HNC3} & 5.2(-11) & 6.4(-12) & C \\ 
\ce{HCCNC} & 3.0(-10) & 8.5(-10) & C & \ce{HC3NH+} & 1.0(-10) & 1.2(-11) & J \\ 
\ce{HCCNCH+} & 3.0(-12) & 1.9(-12) & W & \ce{C3O} & 1.2(-10) & 8.0(-11) & U \\ 
\ce{CH2CHCCH} & 1.2(-09) & 9.1(-13) & H & \ce{HC3O} & 1.3(-11) & 5.8(-10) & U \\ 
\ce{HC3O+} & 2.1(-11) & 8.2(-12) & D & \ce{CH2CHCN} & 6.5(-10) & 5.1(-14) & ZC \\ 
\ce{NCCNH+} & 8.6(-12) & 2.7(-14) & Q & \ce{HCCCHO} & 1.5(-10) & 1.0(-08) & F \\ 
\ce{HCOCN} & 3.5(-11) & 2.4(-10) & F & \ce{C2H5CN} & 1.1(-11) & 1.6(-10) & H \\ 
\ce{CH2CHCHO} & 2.2(-11) & 1.5(-13) & ZJ & \ce{C2S} & 5.5(-09) & 1.6(-11) & A \\ 
\ce{HCCS} & 6.8(-11) & 7.7(-12) & B & \ce{HC2S+} & 1.1(-10) & 7.0(-13) & M \\ 
\ce{NCS} & 7.8(-11) & 1.7(-12) & B & \ce{CH3COCH3} & 1.4(-11) & 1.6(-16) & ZV \\ 
\ce{H2CCS} & 1.8(-10) & 1.8(-14) & B & \ce{HSCN} & 5.8(-11) & 2.4(-13) & B \\ 
\ce{HNCS} & 3.8(-11) & 4.6(-13) & B & \ce{OCS} & <1.8(-09) & 1.2(-10) & ZC \\ 
\ce{HCOOCH3} & 1.1(-10) & 1.5(-16) & ZJ & \ce{C5H} & 1.3(-10) & 3.3(-10) & ZE \\ 
\ce{C5H+} & 8.8(-12) & 1.8(-11) & G & \ce{c-C5H} & 9.0(-12) & 2.1(-12) & ZE \\ 
\ce{C5H2} & 1.4(-12) & 4.0(-11) & K & \ce{c-C3HCCH} & 3.1(-11) & 1.7(-12) & ZB \\ 
\ce{HC4N} & 3.7(-11) & 1.1(-11) & H & \ce{C4O} & <9.0(-12) & 1.4(-08) & U \\ 
\ce{SO2} & 3.0(-10) & 1.1(-13) & ZX & \ce{CH3C4H} & 1.3(-09) & 1.4(-13) & S \\ 
\ce{H2CCCHCCH} & 1.2(-09) & 1.4(-13) & S & \ce{CH2C3N} & 1.6(-11) & 9.1(-11) & N \\ 
\ce{HC4O} & <9.0(-12) & 7.9(-09) & U & \ce{CH3C3N} & 1.7(-10) & 1.6(-10) & P \\ 
\ce{HCCCH2CN} & 2.8(-10) & 4.6(-14) & P & \ce{H2CCCHCN} & 2.7(-10) & 5.3(-13) & P \\ 
\ce{C5H6} & 1.2(-09) & 1.1(-20) & ZB & \ce{CH3CHCHCN} & 1.8(-11) & 6.2(-24) & ZO \\ 
\ce{C3S} & 1.3(-09) & 1.7(-09) & A & \ce{HC3S} & <2.9(-12) & 3.2(-10) & B \\ 
\ce{HC3S+} & 4.0(-11) & 1.1(-11) & B & \ce{H2CCCS} & 7.8(-12) & 6.2(-14) & B \\ 
\ce{HCSC2H} & 3.2(-11) & 4.4(-11) & F & \ce{HCSCN} & 1.3(-10) & 8.9(-13) & F \\ 
\ce{C6H} & 4.8(-10) & 1.5(-10) & YA & \ce{C6H-} & 1.6(-11) & 3.8(-12) & ZK \\ 
\ce{C6H2} & 8.0(-12) & 1.1(-11) & K & \ce{C5N} & 4.7(-11) & 2.3(-11) & YA \\ 
\ce{C5N-} & 8.8(-12) & 1.2(-12) & YA & \ce{HC5N} & 1.8(-08) & 4.7(-09) & C \\ 
\ce{HC4NC} & 3.0(-11) & 1.8(-10) & C & \ce{HC5NH+} & 7.5(-11) & 2.9(-12) & J \\ 
\ce{C6H4} & 5.0(-11) & 2.8(-12) & Y & \ce{C5O} & 1.5(-12) & 1.4(-08) & D \\ 
\ce{HC5O} & 1.4(-10) & 6.4(-09) & U & \ce{NC4NH+} & 1.1(-12) & 1.6(-12) & ZR \\ 
\ce{H2CCHC3N} & 2.0(-11) & 1.2(-14) & ZI & \ce{HCCCHCHCN} & 3.0(-11) & 2.2(-14) & ZI \\ 
\ce{C4S} & 3.8(-12) & 8.2(-10) & B & \ce{HC4S} & 9.5(-12) & 5.7(-13) & ZQ \\ 
\ce{C7H} & 6.5(-12) & 5.1(-11) & ZK & \ce{C6O} & <1.1(-11) & 8.8(-09) & U \\ 
\ce{CH3C6H} & 7.0(-11) & 2.2(-14) & ZF & \ce{H2CCCHC4H} & 2.2(-10) & 1.4(-11) & ZF \\ 
\ce{CH3C5N} & 9.5(-12) & 2.3(-11) & ZF & \ce{H2CCCHC3N} & 1.2(-11) & 1.3(-15) & ZF \\ 
\ce{HC6O} & <1.8(-11) & 4.4(-09) & U & \ce{C5H5CCH} & 3.4(-10) & 9.6(-14) & ZP \\ 
\ce{C5H4CCH2} & 2.7(-10) & 1.6(-13) & ZK & \ce{C5H5CN} & 1.0(-10) & 1.5(-20) & Z \\ 
\ce{C5S} & 5.0(-12) & 1.8(-12) & B & \ce{C8H} & 4.6(-11) & 4.3(-11) & ZH \\ 
\ce{C8H-} & 2.1(-12) & 1.1(-12) & ZH & \ce{C7N-} & 5.0(-12) & 2.9(-13) & ZT \\ 
\ce{HC7N} & 6.4(-09) & 2.4(-10) & C & \ce{HC7NH+} & 5.5(-12) & 6.9(-13) & L \\ 
\ce{C7O} & <2.6(-12) & 6.6(-09) & ZN & \ce{HC7O} & 6.5(-11) & 5.3(-09) & U \\ 
\ce{C6H5CCH} & 2.5(-10) & 6.9(-13) & ZP & \ce{C6H5CN} & 1.6(-10) & 7.6(-13) & ZA \\ 
\ce{C9H} & <3.5(-12) & 1.2(-11) & ZK & \ce{CH3C8H} & <9.8(-10) & 5.9(-21) & V \\ 
\ce{CH3C7N} & 8.6(-12) & 9.7(-13) & V & \ce{C10H-} & 4.0(-11) & 2.8(-13) & ZS \\ 
\ce{C10H} & 2.0(-11) & 1.4(-11) & ZS & \ce{HC9N} & 1.1(-09) & 9.9(-11) & ZC \\ 
\ce{HC11N} & 1.0(-10) & 2.2(-12) & ZD &  &  &  &  \\

\end{longtable}

References: A: \citet{cer21a}; B: \citet{cer21b}; C: \citet{cer20c}; D: \citet{cer20d}; 
E: \citet{cer18e}; F: \citet{cer21f}; G: \citet{cer22g}; H: \citet{cer21h}; 
I: \citet{cer20i}; J: \citet{mar20j}; K: \citet{cab21k}; L: \citet{cab22l}; 
M: \citet{cab22m}; N: \citet{cab21n}; O: \citet{cab21o}; P: \citet{mar21p}: 
Q: \citet{agu15q}; R: \citet{cer21r}; S: \citet{cer21s}; T: \citet{agu22t}; 
U: \citet{cer21u}; V: \citet{sie22v}; W: \citet{agu22w}; Y: \citet{cer21y}; 
Z: \citet{lee21z}; ZA: \citet{bur21za}; ZB: \citet{cer21zb}; ZC: \citet{gra16zc}; 
ZD: \citet{loo21zd}; ZE: \citet{cab22ze}; ZF: \citet{fue22zf}; ZG: \citet{agu13zg}; 
ZH: \citet{bru07zh}; ZI: \citet{lee21zi}; ZJ: \citet{agu21zj}; ZK: \citet{cer22zk}; 
ZL: \citet{loi16zl}; ZM: \citet{cab21zm}; ZN: \citet{cor17zn}; ZO: \citet{cer22zo}; 
ZP: \citet{cer21zp}; ZQ: \citet{fue22zq}; ZR: \citet{agu23zr}; ZS: \citet{rem23zs}; 
ZT: \citet{cer23zt}; ZU: \citet{kal04}; ZV: \citet{agu23zv}; ZW: \citet{pra97}; 
ZX: \citet{cer11}; ZY: \cite{mce13}; ZZ: \citet{mar07}; YA: \citet{agu23ya}; 
YB: \citet{sil23}.

%}
%

\newpage

%  \longtab[2]{
\begin{longtable}{lrrclrrc}
\caption{\label{app:tmc_compar_Crich} Observed fractional abundances, relative to H$_2$, in TMC-1(CP) and corresponding calculated `best fit' values at 1.0 $\times$ 10$^6$ years for the C-rich (C/O = 1.4) model. The `Ref.' column gives references for the observed values. $a(b)=a \times 10^{b}$. }

%\begin{tabular}{lrrclrrc}
\\
\hline\hline
Species   & Observed   &  Calculated  & Ref.   & Species  & Observed  & Calculated  &Ref.  \\
\hline
\endfirsthead
\caption{continued.}\\
\hline\hline
Species   & Observed   &  Calculated  & Ref.   & Species  & Observed  & Calculated  &Ref.  \\
\hline
\endhead
\hline
\endfoot

\ce{OH} & 3.0(-07) & 3.6(-08) & ZG & \ce{NH3} & 5.0(-08) & 4.7(-08) & ZC \\ 
\ce{H2O} & <7.0(-08) & 9.4(-08) & ZG & \ce{C2H} & 6.5(-08) & 6.1(-09) & ZK \\ 
\ce{CN} & 5.0(-09) & 7.2(-08) & ZY & \ce{HCN} & 1.1(-08) & 1.4(-07) & ZW \\ 
\ce{HNC} & 2.6(-08) & 1.8(-07) & ZW & \ce{HCNH+} & 1.9(-09) & 8.9(-10) & ZG \\ 
\ce{H2CN} & 1.5(-11) & 1.0(-11) & ZG & \ce{H2NC} & <3.2(-11) & 4.2(-12) & ZM \\ 
\ce{CO} & 1.7(-04) & 1.9(-04) & ZW & \ce{HCO} & 1.1(-10) & 5.2(-11) & U \\ 
\ce{HCO+} & 9.3(-09) & 5.5(-09) & ZW & \ce{N2H+} & 4.0(-10) & 7.0(-10) & ZY \\ 
\ce{CH2NH} & <3.6(-09) & 3.5(-09) & ZU & \ce{NO} & 3.0(-08) & 5.3(-08) & ZY \\ 
\ce{H2CO} & 5.0(-08) & 5.7(-09) & F & \ce{H2COH+} & <3.0(-11) & 4.1(-12) & D \\ 
\ce{CH3OH} & 4.8(-09) & 2.7(-13) & D & \ce{H2S} & 5.0(-10) & 5.6(-10) & ZY \\ 
\ce{C3H} & 3.0(-09) & 3.3(-09) & ZC & \ce{C3H+} & 2.4(-12) & 4.0(-14) & G \\ 
\ce{C3H2} & 1.9(-09) & 8.3(-09) & ZC & \ce{H2CCC} & 1.9(-10) & 1.4(-11) & K \\ 
\ce{H2CCCH+} & 7.0(-11) & 1.0(-10) & YB & \ce{CH2CCH} & 1.0(-08) & 3.0(-11) & T \\ 
\ce{HCCN} & 4.4(-11) & 7.5(-10) & H & \ce{CH3CCH} & 1.2(-08) & 1.8(-09) & ZC \\ 
\ce{C2O} & 7.5(-11) & 1.1(-09) & U & \ce{CH2CN} & 1.5(-09) & 1.1(-11) & O \\ 
\ce{HCCO} & 7.7(-11) & 7.5(-13) & U & \ce{CH3CN} & 4.7(-10) & 4.6(-10) & O \\ 
\ce{CH2CO} & 1.4(-09) & 7.2(-10) & D & \ce{CH3CHCH2} & 4.0(-09) & 3.4(-16) & ZZ \\ 
\ce{HNCO} & 1.3(-09) & 3.3(-11) & D & \ce{HCNO} & 7.0(-12) & 5.3(-13) & D \\ 
\ce{HOCN} & 1.0(-11) & 5.6(-12) & D & \ce{CH3CO+} & 3.2(-11) & 9.2(-13) & R \\ 
\ce{CS} & 3.5(-08) & 1.1(-07) & B & \ce{CH3CHO} & 3.5(-10) & 5.1(-12) & D \\ 
\ce{CH2CHOH} & 2.5(-10) & 1.4(-11) & ZJ & \ce{H2NCO+} & <4.0(-12) & 3.5(-12) & D \\ 
\ce{HCS} & 5.5(-10) & 2.3(-12) & B & \ce{HCS+} & 1.0(-09) & 6.5(-11) & A \\ 
\ce{NH2CHO} & <5.0(-12) & 0 & D & \ce{HCO2+} & 4.0(-11) & 4.3(-12) & D \\ 
\ce{HCOOH} & 1.4(-10) & 2.5(-11) & D & \ce{CH3OCH3} & 2.5(-10) & 9.2(-19) & ZJ \\ 
\ce{C2H5OH} & 1.1(-10) & 7.0(-13) & ZV & \ce{H2CS} & 4.7(-09) & 1.2(-08) & B \\ 
\ce{NS} & 1.7(-10) & 5.0(-10) & E & \ce{NS+} & 5.2(-12) & 7.3(-12) & E \\ 
\ce{SO} & 1.0(-08) & 3.9(-08) & ZY & \ce{C4H} & 8.5(-09) & 9.4(-09) & YA \\ 
\ce{C4H-} & 2.1(-12) & 6.3(-13) & ZK & \ce{C3N} & 1.2(-09) & 2.1(-09) & YA \\ 
\ce{C3N-} & 6.4(-12) & 6.7(-13) & YA & \ce{C4H2} & 3.3(-10) & 2.3(-09) & K \\ 
\ce{HC3N} & 2.3(-08) & 1.1(-08) & C & \ce{HNC3} & 5.2(-11) & 1.5(-11) & C \\ 
\ce{HCCNC} & 3.0(-10) & 8.2(-10) & C & \ce{HC3NH+} & 1.0(-10) & 9.8(-12) & J \\ 
\ce{HCCNCH+} & 3.0(-12) & 1.6(-12) & W & \ce{C3O} & 1.2(-10) & 8.9(-10) & U \\ 
\ce{CH2CHCCH} & 1.2(-09) & 6.6(-10) & H & \ce{HC3O} & 1.3(-11) & 5.3(-11) & U \\ 
\ce{HC3O+} & 2.1(-11) & 2.8(-12) & D & \ce{CH2CHCN} & 6.5(-10) & 7.3(-10) & ZC \\ 
\ce{NCCNH+} & 8.6(-12) & 1.4(-11) & Q & \ce{HCCCHO} & 1.5(-10) & 2.5(-10) & F \\ 
\ce{HCOCN} & 3.5(-11) & 6.8(-11) & F & \ce{C2H5CN} & 1.1(-11) & 3.0(-11) & H \\ 
\ce{CH2CHCHO} & 2.2(-11) & 1.6(-13) & ZJ & \ce{C2S} & 5.5(-09) & 5.1(-09) & A \\ 
\ce{HCCS} & 6.8(-11) & 1.1(-10) & B & \ce{HC2S+} & 1.1(-10) & 1.3(-11) & M \\ 
\ce{NCS} & 7.8(-11) & 7.2(-11) & B & \ce{CH3COCH3} & 1.4(-11) & 2.8(-17) & ZV \\ 
\ce{H2CCS} & 1.8(-10) & 4.9(-11) & B & \ce{HSCN} & 5.8(-11) & 1.9(-11) & B \\ 
\ce{HNCS} & 3.8(-11) & 3.3(-11) & B & \ce{OCS} & <1.8(-09) & 2.7(-08) & ZC \\ 
\ce{HCOOCH3} & 1.1(-10) & 8.0(-19) & ZJ & \ce{C5H} & 1.3(-10) & 5.3(-09) & ZE \\ 
\ce{C5H+} & 8.8(-12) & 1.8(-11) & G & \ce{c-C5H} & 9.0(-12) & 5.3(-09) & ZE \\ 
\ce{C5H2} & 1.4(-12) & 1.4(-07) & K & \ce{c-C3HCCH} & 3.1(-11) & 4.8(-11) & ZB \\ 
\ce{HC4N} & 3.7(-11) & 1.3(-09) & H & \ce{C4O} & <9.0(-12) & 2.2(-09) & U \\ 
\ce{SO2} & 3.0(-10) & 2.8(-09) & ZX & \ce{CH3C4H} & 1.3(-09) & 6.1(-11) & S \\ 
\ce{H2CCCHCCH} & 1.2(-09) & 1.4(-10) & S & \ce{CH2C3N} & 1.6(-11) & 1.8(-11) & N \\ 
\ce{HC4O} & <9.0(-12) & 3.4(-09) & U & \ce{CH3C3N} & 1.7(-10) & 1.2(-09) & P \\ 
\ce{HCCCH2CN} & 2.8(-10) & 8.2(-11) & P & \ce{H2CCCHCN} & 2.7(-10) & 8.8(-10) & P \\ 
\ce{C5H6} & 1.2(-09) & 6.8(-15) & ZB & \ce{CH3CHCHCN} & 1.8(-11) & 2.4(-17) & ZO \\ 
\ce{C3S} & 1.3(-09) & 9.4(-09) & A & \ce{HC3S} & <2.4(-11) & 1.4(-11) & B \\ 
\ce{HC3S+} & 4.0(-11) & 4.7(-11) & B & \ce{H2CCCS} & 7.8(-12) & 4.7(-11) & B \\ 
\ce{HCSC2H} & 3.2(-11) & 2.3(-10) & F & \ce{HCSCN} & 1.3(-10) & 7.0(-10) & F \\ 
\ce{C6H} & 4.8(-10) & 6.3(-09) & YA & \ce{C6H-} & 1.6(-11) & 2.6(-10) & ZK \\ 
\ce{C6H2} & 8.0(-12) & 5.1(-07) & K & \ce{C5N} & 4.7(-11) & 9.8(-11) & YA \\ 
\ce{C5N-} & 8.8(-12) & 7.2(-12) & YA & \ce{HC5N} & 1.8(-08) & 2.5(-08) & C \\ 
\ce{HC4NC} & 3.0(-11) & 4.8(-10) & C & \ce{HC5NH+} & 7.5(-11) & 1.0(-11) & J \\ 
\ce{C6H4} & 5.0(-11) & 7.3(-10) & Y & \ce{C5O} & 1.5(-12) & 1.3(-09) & D \\ 
\ce{HC5O} & 1.4(-10) & 1.7(-09) & U & \ce{NC4NH+} & 1.1(-12) & 5.2(-11) & ZR \\ 
\ce{H2CCHC3N} & 2.0(-11) & 7.2(-11) & ZI & \ce{HCCCHCHCN} & 3.0(-11) & 1.4(-10) & ZI \\ 
\ce{C4S} & 3.8(-12) & 5.3(-09) & B & \ce{HC4S} & 9.5(-12) & 1.1(-09) & ZQ \\ 
\ce{C7H} & 6.5(-12) & 2.5(-09) & ZK & \ce{C6O} & <1.1(-11) & 1.4(-08) & U \\ 
\ce{CH3C6H} & 7.0(-11) & 8.0(-11) & ZF & \ce{H2CCCHC4H} & 2.2(-10) & 1.1(-10) & ZF \\ 
\ce{CH3C5N} & 9.5(-12) & 1.1(-11) & ZF & \ce{H2CCCHC3N} & 1.2(-11) & 8.9(-12) & ZF \\ 
\ce{HC6O} & <1.8(-11) & 6.2(-09) & U & \ce{C5H5CCH} & 3.4(-10) & 4.1(-11) & ZP \\ 
\ce{C5H4CCH2} & 2.7(-10) & 2.7(-10) & ZK & \ce{C5H5CN} & 1.0(-10) & 2.5(-15) & Z \\ 
\ce{C5S} & 5.0(-12) & 2.4(-09) & B & \ce{C8H} & 4.6(-11) & 7.2(-09) & ZH \\ 
\ce{C8H-} & 2.1(-12) & 1.7(-10) & ZH & \ce{C7N-} & 5.0(-12) & 2.0(-10) & ZT \\ 
\ce{HC7N} & 6.4(-09) & 4.0(-07) & C & \ce{HC7NH+} & 5.5(-12) & 6.5(-10) & L \\ 
\ce{C7O} & <2.6(-12) & 1.3(-07) & ZN & \ce{HC7O} & 6.5(-11) & 1.2(-07) & U \\ 
\ce{C6H5CCH} & 2.5(-10) & 3.4(-10) & ZP & \ce{C6H5CN} & 1.6(-10) & 2.0(-10) & ZA \\ 
\ce{C9H} & <3.5(-12) & 3.4(-09) & ZK & \ce{CH3C8H} & <9.8(-10) & 3.0(-12) & V \\ 
\ce{CH3C7N} & 8.6(-12) & 1.1(-10) & V & \ce{C10H-} & 4.0(-11) & 1.1(-10) & ZS \\ 
\ce{C10H} & 2.0(-11) & 4.5(-09) & ZS & \ce{HC9N} & 1.1(-09) & 5.0(-07) & ZC \\ 
\ce{HC11N} & 1.0(-10) & 1.7(-07) & ZD &  &  &  &  \\

\end{longtable}

References: A: \citet{cer21a}; B: \citet{cer21b}; C: \citet{cer20c}; D: \citet{cer20d}; 
E: \citet{cer18e}; F: \citet{cer21f}; G: \citet{cer22g}; H: \citet{cer21h}; I: \citet{cer20i}; 
J: \citet{mar20j}; K: \citet{cab21k}; L: \citet{cab22l}; M: \citet{cab22m}; N: \citet{cab21n}; 
O: \citet{cab21o}; P: \citet{mar21p}: Q: \citet{agu15q}; R: \citet{cer21r}; S: \citet{cer21s}; 
T: \citet{agu22t}; U: \citet{cer21u}; V: \citet{sie22v}; W: \citet{agu22w}; Y: \citet{cer21y}; 
Z: \citet{lee21z}; ZA: \citet{bur21za}; ZB: \citet{cer21zb}; ZC: \citet{gra16zc}; ZD: \citet{loo21zd}; 
ZE: \citet{cab22ze}; ZF: \citet{fue22zf}; ZG: \citet{agu13zg}; ZH: \citet{bru07zh}; 
ZI: \citet{lee21zi}; ZJ: \citet{agu21zj}; ZK: \citet{cer22zk}; ZL: \citet{loi16zl}; 
ZM: \citet{cab21zm}; ZN: \citet{cor17zn}; ZO: \citet{cer22zo}; ZP: \citet{cer21zp}; 
ZQ: \citet{fue22zq}; ZR: \citet{agu23zr}; ZS: \citet{rem23zs}; ZT: \citet{cer23zt}; 
ZU: \citet{kal04}; ZV: \citet{agu23zv}; ZW: \citet{pra97}; ZX: \citet{cer11}; ZY: \cite{mce13}; 
ZZ: \citet{mar07}; YA: \citet{agu23ya}; YB: \citet{sil23}.

%}
%

\end{appendix}

\end{document}